\newcommand{\La}{\mathcal{L}}
\newcommand{\tr}{\text{tr}}

\newcommand{\ZZp}{\mathbf{Z}'}
\newcommand{\ZZ}{\mathbf{Z}}
\newcommand{\RR}{\mathbf{R}}
\newcommand{\RRp}{\mathbf{R}'}
\newcommand{\rom}{\mathrm}
\newcommand{\LL}{\mathbf{L}}
\newcommand{\LLp}{\mathbf{L}'}

\newcommand{\ket}[1]{{\left|#1\right\rangle}}
\newcommand{\bra}[1]{{\left\langle #1\right|}}

\documentclass{iopjournal}

\usepackage{float}
\usepackage{tikz}
\usepackage{orcidlink}
\usepackage{amsmath}
\usepackage{ragged2e}
\usepackage{microtype}
\usepackage[bitstream-charter]{mathdesign}
\usetikzlibrary{decorations.pathreplacing} 
\usetikzlibrary{positioning}
\usetikzlibrary{calc}
\usepackage{tcolorbox}
\usepackage{booktabs}
\begin{document}
\pagestyle{plain}
\justifying
\title{On the integrability structure of the deformed rule-54 reversible cellular automaton}

\author{Chiara Paletta$^{1}$\orcid{0000-0003-3660-9806}, Toma\v{z} Prosen$^{1,2}$\orcid{0000-0001-9979-6253}}

\affil{$^1$Department of Physics, Faculty of Mathematics and Physics,
 University of Ljubljana, Jadranska 19, SI-1000 Ljubljana, Slovenia\\}

\affil{$^2$Institute of Mathematics, Physics and Mechanics, Jadranska 19, SI-1000, Ljubljana, Slovenia}

\email{chiara.paletta@fmf.uni-lj.si, tomaz.prosen@fmf.uni-lj.si}


\begin{abstract}
{
We study quantum and stochastic deformations of the rule-54 reversible cellular automaton (RCA54) on a 1+1--dimensional spatiotemporal lattice, focusing on their integrability structures in two distinct settings. First, for the quantum deformation, which turns the model into an interaction-round-a-face brickwork quantum circuit (either on an infinite lattice or with periodic boundary conditions), we show that the shortest-range nontrivial conserved charge commuting with the discrete-time evolution operator has a density supported on six consecutive sites. By constructing the corresponding range-6 Lax operator, we prove that this charge belongs to an infinite tower of mutually commuting conserved charges generated by higher-order logarithmic derivatives of the transfer matrix. With the aid of an intertwining operator, we further prove that the transfer matrix commutes with the discrete-time evolution operator.

Second, for the stochastic deformation, which renders the model into a Markov-chain circuit, we investigate open boundary conditions that couple the system at its edges to stochastic reservoirs. In this setting, we explicitly construct the non-equilibrium steady state (NESS) by means of a staggered patch matrix ansatz, a hybrid construction combining the previously used commutative patch-state ansatz for the undeformed RCA54 with the matrix-product ansatz. Finally, we propose a simple empirical criterion for detecting integrability or exact solvability in a given model setup, introducing the notion of digit complexity.
}
\end{abstract}

\tableofcontents

\section{Introduction}

The theory of integrable interacting particle systems is a multifaceted area of mathematical physics that lies at the intersection of three physically distinct domains: (i) integrable deterministic classical dynamical systems~\cite{faddeev/takhtajan1,babelonbernardtalon}, (ii) integrable quantum systems~\cite{korepinbogoliubovizergin,faddeev1996algebraic}, and (iii) integrable classical stochastic systems~\cite{schutz2001}. From the perspective of physical applications, for example in transport theory, domains (i) and (ii) are more closely related, since both describe dissipationless, conservative, i.e.\ Hamiltonian, dynamics, in classical and quantum settings, respectively. By contrast, domains (ii) and (iii) are closer in terms of mathematical methodology, as both rely on variants of the algebraic Bethe ansatz, in real and imaginary time, respectively, whereas domain (i) is rooted in classical soliton theory and the nonlinear inverse scattering transform.

Whereas classical deterministic integrable systems are usually formulated in terms of continuous phase-space variables, quantum and stochastic models are typically defined on a Hilbert space spanned by a discrete set of configurations. A fruitful paradigm that bridges all three domains is provided by reversible cellular automata, whose equations of motion are specified by reversible local update rules. Yet, whereas the integrability structures of classical lattice systems with continuous variables, such as the Toda lattice, are well understood, the situation for reversible cellular automata is very different. Although such automata were classified long ago~\cite{takesue1987,bobenko1993two}, their integrability has not been studied systematically. Reversible cellular automaton rule 54 (RCA54) has been identified as a possibly minimal model of classical soliton theory~\cite{bobenko1993two}. Moreover, while many exact solutions for nonequilibrium statistical-mechanics setups of this model have appeared in recent years~\cite{buvca2021rule}, its connection to canonical integrability structures---such as Lax operators and Yang--Baxter equations---has only recently begun to emerge~\cite{prosen2021reversible,gombor2024integrable}, and most fundamental questions remain open.

RCA54 may be viewed as perhaps the simplest microscopic deterministic theory in $1+1$ dimensions that combines strong local interactions with asymptotically free propagation of excitations, namely quasiparticles or solitons. In this minimal model, not only space and time but also the field variables themselves are discrete: each cell takes only two values, $0$ or $1$. The model is most naturally written in light-cone coordinates, which mimic a relativistically covariant theory in a fully discrete setting. These simple algebraic structures have enabled not only exact but fully explicit calculations of a variety of nonequilibrium and dynamical quantities, including the dynamical structure factor~\cite{klobas2019time}, relaxation after an inhomogeneous quench~\cite{klobas2019time}, the steady state~\cite{prosen2016integrability}, and even the full diagonalization of the corresponding stochastically boundary-driven cellular chain~\cite{prosen2017} (see Ref.~\cite{buvca2021rule} for a review). The model has also been studied in a quantum setting~\cite{gopalakrishnan2018facilitated}, where its simple yet nontrivial dynamics provide an ideal framework for investigating large-scale behavior of local observables and operator spreading~\cite{alba2019,piroli2021}. In addition, RCA54 admits an interpretation analogous to a spin-glass model, as a discrete-time deterministic counterpart of the Fredrickson--Andersen model~\cite{fredrickson1984kinetic}.

At the same time, RCA54 is a highly degenerate integrable---or, more precisely, superintegrable---model, in the sense that the number of local conserved charges grows exponentially with their support size. In \cite{Friedman19}, a deformation preserving integrability was introduced, which gives dispersion to the quasiparticles. In particular, a translationally invariant range-6 charge commuting with the circuit time evolution was found. The simple structure of the model also allowed one to derive and solve the Bethe equations. This work hinted at a possible Yang–Baxter integrability structure of the model, which was later clarified in \cite{gombor2024integrable}. There, a quantum Lax operator was constructed, generating a family of conserved charges that commute with the discrete-time dynamics. However, the dynamical map itself could not be recovered from that Lax operator or from the corresponding transfer matrix.

When viewed as a two-dimensional model, RCA54 is naturally formulated as an interaction-round-a-face (IRF) model~\cite{baxter2016exactly} rather than as a more conventional vertex model. This raises the question of whether RCA54 admits simple one-parameter or multi-parameter deformations in the direction of quantum or statistical/stochastic dynamics. In analogy with the way the six-vertex model maps either to the quantum XXZ circuit~\cite{vanicat2018integrable2} or to a parallel-update simple exclusion process~\cite{vanicat2018integrable}, one may ask whether RCA54 is merely the deterministic skeleton of a richer family of integrable---though no longer superintegrable---quantum or stochastic dynamics.

This is precisely the direction pursued in the present work. We study a quantum or stochastic IRF circuit that may be viewed as being composed of generalized Toffoli (three-site control-control) gates, in which the middle qubit (cell) is updated according to the states of its left and right neighbors. In deterministic RCA54, the middle cell is flipped whenever at least one of its neighbors is in state $1$. Here we allow two additional processes: when both neighbors are in state $0$, we apply a nontrivial quantum or stochastic map to the middle qubit/cell. In the stochastic setting, this amounts to a single-bit Markov chain acting on the middle cell, thereby allowing particle pair-creation and pair-annihilation processes. Figure~\ref{figuretraj} shows a typical trajectory of the resulting stochastic cellular automaton. This stochastic model appears to provide a doubly space-time-discretized version of the so-called $t$-PNG model, a $q$-deformed polynuclear growth model known to belong to the Kardar--Parisi--Zhang universality class~\cite{borodin2023}. Preliminary investigations of the stochastically deformed RCA54 also indicate KPZ scaling, although the details will be discussed elsewhere.

In this work, we focus on uncovering Yang--Baxter-type integrability structures in the deformed RCA54 model. Our main findings are summarized below.

\subsection{Summary of the results}

In what follows, we study the integrability of quantum and stochastic deformations of the RCA54 model. In Sec.~\ref{dynamics}, we first describe the dynamics of the deformed model. For periodic boundary conditions, the most general deformation is parametrized by four complex parameters $(\alpha, \beta, \gamma, \delta)$. For open boundary conditions, by contrast, the dynamics must be stochastic, with stochastic baths coupled to the edges, and only two real parameters $(\beta, \gamma \in [0,1])$ remain relevant. Under \textbf{periodic boundary conditions}, \textit{integrability} refers to an integrable structure of the full spectrum, meaning that the propagator $\mathbb{U}$ commutes with an infinite family of conserved charges. Under \textbf{open boundary conditions}, it refers instead\footnote{We remark that, based on numerical analysis of spectral statistics, the spectrum of the corresponding many-particle Markov chain also shows convincing signatures of integrability, using the method of Ref.~\cite{sa2020complex}. A similar integrability conjecture appears to hold for a boundary-driven XXZ quantum circuit with boundary reset gates~\cite{popkov2025exact}, where, in contrast to Ref.~\cite{paletta2025integrability}, the boundary gates do not satisfy the reflection-algebra equations. An analytical proof of these conjectural statements, however, is still lacking.} to the existence of an explicit analytical expression for the non-equilibrium steady state.

We divide the paper into two conceptually related but technically largely independent parts:

\begin{itemize}
\item \textit{Closed system on an infinite lattice or with periodic boundary conditions.} In Sec.~\ref{bulksection}, we show that the discrete-time evolution operator $\mathbb{U}$ of the quantum deformed RCA54 model commutes with a range-6 operator $Q_6$ that is invariant under translations by two sites. This operator plays a central role throughout our analysis and serves as the basic object from which the integrability of the model is established.

In Sec.~\ref{recaponint}, we briefly review the formalism of medium-range\footnote{The term \textit{medium} refers to systems whose interaction range is greater than two sites (i.e.\ beyond nearest neighbor) but still finite, in contrast to genuinely long-range models.} Yang--Baxter integrability introduced in \cite{sajat-medium}. Sections~\ref{frameworkint} and \ref{evidenceofintegrability} contain the main new results of the paper: Sec.~\ref{frameworkint} gives an overview of the central ideas and conclusions, while Sec.~\ref{evidenceofintegrability} provides the full technical development. In Sec.~\ref{perturbativesketch}, we show how the deformed RCA54 model can be embedded into the framework of Yang--Baxter integrability for arbitrary values of the four deformation parameters. In particular, we demonstrate that the range-6 charge belongs to a hierarchy of commuting charges. Starting from the formal construction of commuting charges via logarithmic derivatives of the transfer matrix, we show how two additional charges can be generated from the range-6 operator. This approach is conceptually similar to those of Refs.~\cite{de2023lifting,Gombor:2022lco}, where one tests whether a nearest-neighbor (n.n.) Hamiltonian belongs to a Yang--Baxter-integrable tower of conserved charges.

For nearest-neighbor models, higher commuting charges can be generated from the boost operator, or equivalently from logarithmic derivatives of the transfer matrix. For medium-range models, however, no analogue of the boost operator is presently available, so one must work directly with the transfer-matrix construction. Recently, Ref.~\cite{hokkyo2025integrability} proved the star--triangle conjecture for nearest-neighbor models: the Reshetikhin condition ensuring commutativity of the first two charges is both necessary and sufficient for the existence of an infinite family of commuting charges. Moreover, Ref.~\cite{zhang2026bootstrapping} provided a program to bootstrap the $R$-matrix starting from the first charge. In Sec.~\ref{perturbativesketch}, we show how the range-6 operator can be recast in nearest-neighbor form by gluing, or fusing, pairs of adjacent sites twice. Combined with the result of Ref.~\cite{hokkyo2025integrability}, this allows us to infer the existence of a tower of mutually commuting charges. We obtain analytical expressions for the next two conserved operators, $Q_{10}$ and $Q_{14}$, of ranges 10 and 14, respectively, and verify that they commute with the discrete-time evolution operator of the deformed RCA54 model. A full proof of Yang--Baxter integrability for the deformed RCA54 model would imply that all such charges are conserved, and hence commute with the discrete-time evolution operator for arbitrary deformation parameters. At present, we have established this explicitly for the first three charges only. These three charges are then used to construct a perturbative Lax operator, which serves as the basic building block of the transfer matrix.

\begin{figure}[H]
    \centering    \includegraphics[width=0.6\textwidth,angle=45]{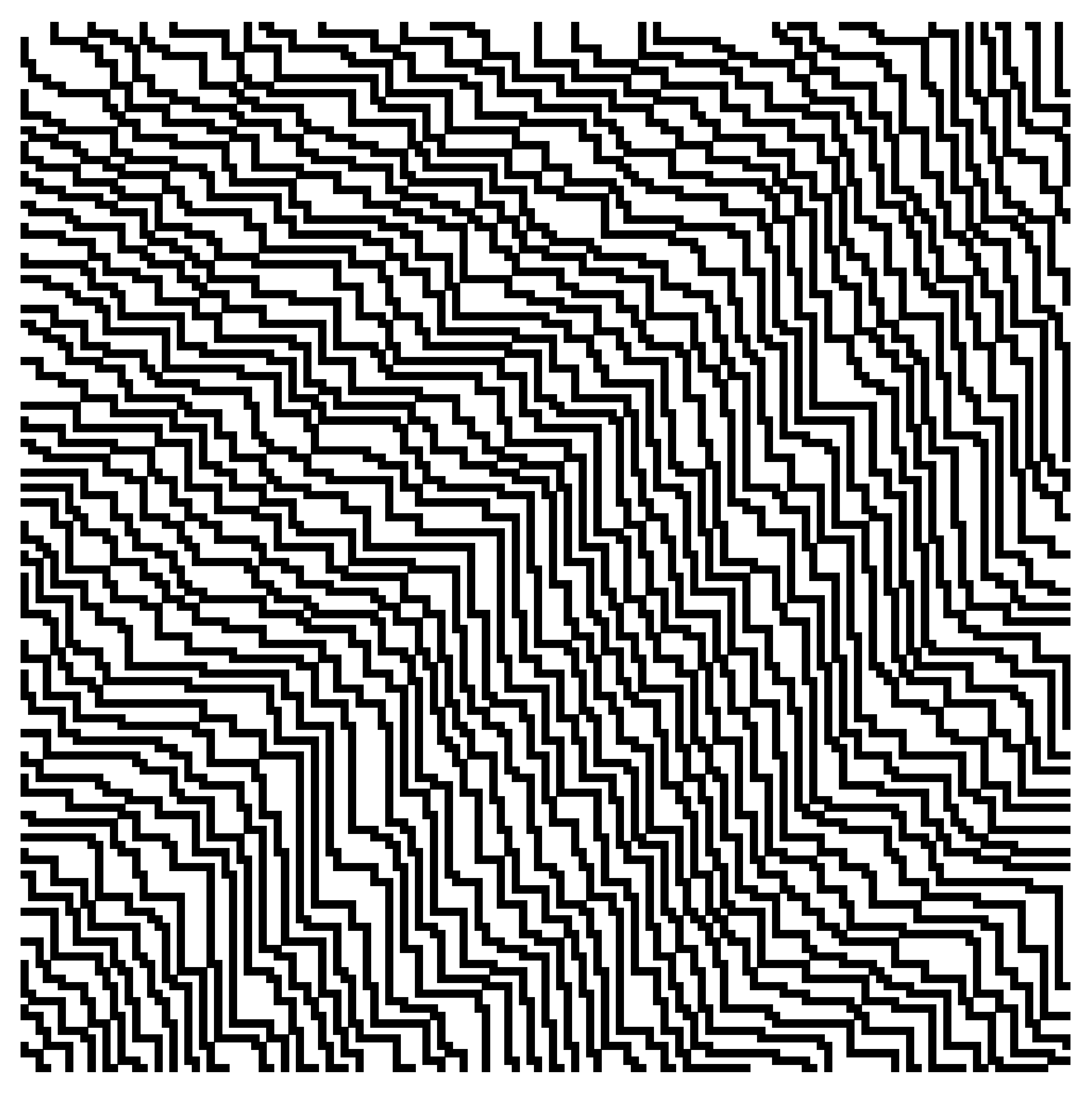}
    
\includegraphics[width=0.83\textwidth]{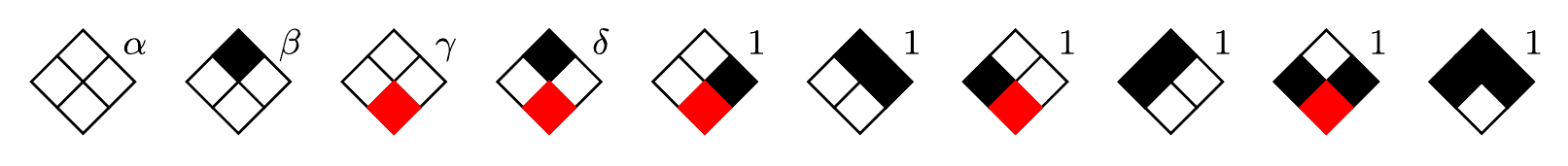}
    
    \caption{A typical trajectory of stochastically deformed RCA54 automaton with random `light-cone' initial condition. Here time runs top-down. The stochastic rule is encoded in terms of 
    10 face plaquettes (below), with probability indicated on top-right of each diagram, and where dynamically updated cells (empty$=0$, filled$=1$) are indicated in red. Here: $\alpha=1-\gamma=0.2$, $\delta=1-\beta=0.8$.}
    \label{figuretraj}
\end{figure}

In the accompanying Mathematica notebook \texttt{BulkIntegrability.nb}~\cite{mathematicanotebook}, we provide analytical expressions for the charges $Q_6$, $Q_{10}$, and $Q_{14}$, together with the perturbative expansion of the Lax operator. The organization of the notebook is described in Appendix~\ref{notebookappendix}. In Sec.~\ref{analyticsketch}, we derive a symbolic nonperturbative expression for the Lax operator at a generic fixed choice of deformation parameters. Our algorithm can in principle be applied to any fixed choice of parameters, although a fully analytic closed-form expression valid for variable parameters remains beyond the reach of the present method. The nonzero entries of the Lax operator are displayed in Fig.~\ref{nonzeroLax}, and the complete expression is included in Ref.~\cite{mathematicanotebook} (\texttt{BulkIntegrability.nb}). In Sec.~\ref{prooftutv}, we verify the existence of the intertwining operator, i.e.\ the $R$-matrix ensuring commutativity of transfer matrices at different values of the spectral parameter. Finally, in Sec.~\ref{prooftuU}, we prove that the transfer matrix commutes with the deformed RCA54 evolution operator by establishing the existence of a second intertwining operator, $A$, which guarantees the time independence of all commuting charges. Intuitive graphical arguments in \emph{face-tensor} notation are presented in Figs.~\ref{Rllproof} and \ref{tuproof}.

Section~\ref{evidenceofintegrability} contains the full technical details of the program outlined in Sec.~\ref{frameworkint}: a perturbative proof of integrability is given in Sec.~\ref{evidenceofintegrability1}, while the exact analytic construction of the Lax operator is developed in Sec.~\ref{technicaldetailsLax}. In Sec.~\ref{algebraicprooftutv}, we present an algebraic proof of the commutativity of the transfer matrices, and in Sec.~\ref{algebraicprooftuU}, an algebraic proof of the conservation of the charges.

\begin{table}[h]
\centering
\renewcommand{\arraystretch}{1.2}
\begin{tabular}{ll}
\toprule
\multicolumn{2}{c}{\textbf{Summary: Bulk integrability}} \\
\midrule
{Variable $\alpha,\beta,\gamma,\delta$} & \\[2pt]

& Main fact: $[\mathbb{U},Q_6]=0$ \\

& Proof that $[Q_6,Q_i]=0$, \quad $i=4\mathbb{N}+2$ \\

& Analytical expressions of $Q_6,Q_{10},Q_{14}$ \\

& Check $[\mathbb{U},Q_{10}]=[\mathbb{U},Q_{14}]=0$ \\

& Perturbative expression of the Lax operator $\mathcal{L}(u)$ up to $O(u^4)$\\
[6pt]

{Fixed (generic) choice of $\alpha,\beta,\gamma,\delta$} & \\[2pt]

& Analytical expression of the Lax operator $\mathcal{L}(u)$: 
$\partial_u \log t(u)|_{u=0}=Q_6$ \\

& Existence of $R$-operator: RLL relation $\Rightarrow [t(u),t(v)]=0$ \\

& Existence of additional intertwining operator $A$ proving: 
$[t(u),\mathbb{U}]=0$ \\
\bottomrule
\end{tabular}
\end{table}

\item \textit{Open system with stochastic boundaries.} In Sec. \ref{boundarydrivness}, we consider a specific case of a stochastic deformation of RCA54 in the bulk, with the left and right boundaries coupled to stochastic reservoirs of quasi-particles (left and right movers). In this context, we define integrability as exact (or efficient) solvability, i.e. as the ability to analytically  (or efficiently, i.e., in time polynomial in the length of the system, and poly-logarithmic in precision --- or polynomial in the number of accurate digits of the result) determine the non-equilibrium steady state (NESS), i.e. a fixed point of the interacting markovian dynamics.

In Sec. \ref{nesssection}, we define the NESS and in \ref{uniqueness}, we generalize the proof from Ref.~\cite{prosen2016integrability} to establish the existence and uniqueness of the NESS and the relaxation toward it in the setting of stochastic deformation. Then, in Sec. \ref{patch}, we provide an explicit expression for the NESS using a staggered patch matrix product ansatz. This solution involves four boundary vectors ($\LL,\LLp, \RR, \RRp$) and two bulk operators ($\ZZ, \ZZp$) which admit an infinitely-dimensional block-tridiagonal representation with $3\times 3$ blocks. We give a sketch of this tensor network structure in Fig. \ref{figureZansatz}, while in the appendix \ref{masksappendix}, we give a characterization of which entries of the tensors are non-zero. In \cite{mathematicanotebook} (\texttt{BoundaryNESS.nb}), by letting the user choose the deformation parameters, we provide an efficient algorithm for the explicit solution of the boundary vectors and, for the bulk, a recursive construction of the block tridiagonal representation. The boundary operators obey the face version of the Ghoshal-Zamolodchikov relations, while the bulk operators satisfy the corresponding face analog of the Zamolodchikov-Faddeev algebra.
In \cite{mathematicanotebook}, we also provide  an analytical expression in terms of boundary and bulk parameters for boundary vectors $\LL,\LLp, \RR, \RRp$ and the first two matrix blocks of $\ZZ$ and $\ZZp$. We furthermore provide some extra relations among entries of the bulk tensors and explicitly verify our ansatz for generic, fixed choices of parameters up to the 50th step (level) in the recursive construction.

Finally, in Sec. \ref{digitcomplexity}, we propose a possible empirical check of integrability or exact solvability based on the idea of digit complexity, namely for a random rational choice of model parameters, the number $\#(N)$ of digits of the denominator expressing a given test-observable can scale polynomially or exponentially with the size $N$ of the model. 
Benchmarking this method on several other models, including integrable and non-integrable versions of boundary driven stochastic six-vertex model, we conclude that deformed RCA54 is a rather complex integrable model ($\#(N)=O(N^2)$) compared to the integrable stochastic six vertex model ($\#(N)=O(N)$).

\end{itemize}
\section{Deformed Rule 54 Dynamics}
\label{dynamics}
\subsection{Closed system -- periodic boundary condition}

We define the discrete time evolution of the initial state vector $\mathbf{p}(0) \in (\mathbb C^2)^{\otimes N}$, as
\begin{equation}
    \mathbf{p}(t)=\mathbb{U}^t \mathbf{p}(0),
\end{equation}
where the time propagator $\mathbb{U}$ factorizes as
\begin{align}
&\mathbb{U}=\mathbb{U}_\mathrm{e} \mathbb{U}_\mathrm{o},\label{Ucircuit}\\&\mathbb{U}_\mathrm{e}=U_{123}U_{345}\dots U_{N-3,N-2,N-1}U_{N-1,N,1},\label{Ueevolutionperiodic}\\&\mathbb{U}_\mathrm{o}=U_{234}U_{456}\dots U_{N-2,N-1,N}U_{N,1,2},\label{Uoevolutionperiodic}
\end{align}
with $\mathbb{U}, \mathbb{U}_\mathrm{o}, \mathbb{U}_\mathrm{e}\,\in\, \text{End}\big((\mathbb{C}^2)^{\otimes N}\big)$; $N$ is the number of qubits (spins, two-level systems, etc), $U$ is an 8$\times$8 unitary matrix\footnote{Depending on the physics application, $U$ can also be a stochastic or arbitrary/structureless complex matrix.} which encodes the deformed rule 54 (see below), and the indices $({x-1,x,x+1})$ of $U_{x-1,x,x+1}$ identify the sites on which $U$ acts non-trivially. 
Each site variable (cell/spin/bit) can take two possible values, which we identify as $\{0,1\}$. We depict them as empty/white (state $0$) or filled/black (state $1$) boxes (see illustration in Fig.~\ref{figuretraj} for the stochastic implementation). For simplicity, we assume the number of qubits $N$ to be {\em even}.
\begin{figure}[H]
    \centering    \includegraphics[width=0.36\textwidth]{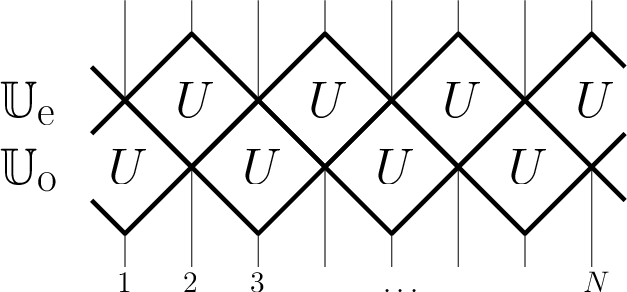}

    \caption{Graphical representation of the dynamics \eqref{Ucircuit}. Time runs bottom-up.}
    \label{figurematrices2}
\end{figure}
The site $x$ in the triple of states $(x-1,x,x+1)$ identifies the active bit; it is modified by the action of the operator $U$ according to the values of the control bits $x-1$ and $x+1$
\begin{align}
    &U_{x-1,x,x+1}=\sum_{i,j,k,l=0}^1(f_{k,l})_i^j 
    \ket{k}_{x-1}\ket{i}_x\ket{l}_{x+1}\bra{k}_{x-1}\bra{j}_x\bra{l}_{x+1},
    \label{propagatorU}
\end{align}
with $f_{k,l}\,\in \,\text{End}(\mathbb{C}^2),\,\,k,l=0,1\,$ representing face-weights as a quadruple of $2\times 2$ matrices
\vspace{-0.5cm}

\begin{figure}[H]
    \centering    \includegraphics[width=0.2\textwidth]{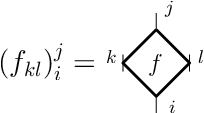}
\end{figure}

\vspace{-0.5cm}
We label the two steps of the evolution operator $\mathbb{U}$ in \eqref{Ucircuit} as even ($\mathbb{U}_\mathrm{e}$) or odd ($\mathbb{U}_\mathrm{o}$) depending on whether active sites are placed on even or odd sites, respectively. The matrices $f_{k,l}$ uniquely determine the quantum (or stochastic) automaton.
For instance, the undeformed RCA54 model is encoded by:
\begin{align}
    f_{00}=\mathbb{1},
    \qquad f_{01}=f_{10}=f_{11}=X,
\end{align}
where $X=\begin{pmatrix} 0 & 1 \cr 1 & 0\end{pmatrix}$ is the Pauli matrix.
The undeformed model is clearly deterministic as $U$ corresponds to a permutation over $S_8$, and consequently $\mathbb{U}$ maps basis (product) states $\ket{i_1,i_2,\dots, i_N}$, $i_x \in {0,1}$ to 
basis states.

The deformed version of RCA54 was introduced in Ref.~\cite{prosen2021reversible}. The dynamics is defined by introducing four deformation parameters $\alpha, \beta, \gamma, \delta \in \,\mathbb{C}$,
\begin{align}
    f_{00}=\left(
\begin{array}{cc}
 \alpha  & \beta  \\
 \gamma  & \delta  \\
\end{array}
\right),
    \qquad f_{01}=f_{10}=f_{11}=X.
\label{deformation}
\end{align}
Hence, the source of non-deterministic property emerges if the control (neighbouring) sites are both in the empty state $0$, explicitly
\begin{align}
    &\ket{000}\to \alpha\ket{000}+\beta\ket{010},\\
    &\ket{010}\to \gamma\ket{000}+\delta\ket{010}.
\end{align}

The dynamics then exhibits different physical paradigms depending on the values of the parameters $\alpha,\beta,\gamma,\delta$:
\begin{itemize}
    \item Deterministic RCA54 $\alpha=\delta=1$ and $\beta=\gamma=0$;
    \item Deterministic trivial dynamics $\beta=\gamma=1$ and $\alpha=\delta=0$ ($f_{00}=X$);
    \item $f_{00}\,\in\,U(2)$ is unitary, the evolution $\mathbb{U}$ corresponds to a local Floquet circuit or a quantum cellular automaton (QCA);
    \item $f_{00}$ is a stochastic matrix, $\alpha,\beta,\gamma,\delta\ge0$ and $\alpha+\gamma=\beta+\delta=1$. $\mathbb{U}$ corresponds to a local Markov chain circuit or stochastic cellular automaton (SCA).
\end{itemize}

For arbitrary values of the deformation parameters, the dynamics preserves a diagonal U(1) charge that can be interpreted as the net soliton current, specifically, the number of right movers minus the number of left movers \cite{prosen2021reversible}
\begin{align}
    &\mathcal{J}=\sum_{j=1}^{N/2} \big(Z_{2j-1}Z_{2j}-Z_{2j}Z_{2j+1}\big),\qquad [\mathbb{U},\mathcal{J}]=0,
    \label{U1likecharge}
\end{align}
where $Z=\begin{pmatrix} 1 & 0 \cr 0 & -1\end{pmatrix}$ is the diagonal Pauli matrix.

In the following, in \ref{perturbativesketch} we show evidence of the algebraic Yang-Baxter integrability of deformed RCA54 for arbitrary deformation parameters. In \ref{analyticsketch}, we prove algebraic integrability for a generic fixed choice of parameters. Our algorithm can, in principle, also be applied to free (variable) parameters; however, the resulting analytical expressions turn out to be prohibitively complicated.

\subsection{Open boundary conditions: stochastic boundary driving}
\label{openboundaryconditiondynamics}
We now restrict to the case where $f_{00}$ is a stochastic matrix, hence the evolution in the bulk is given by a stochastic cellular automaton
\begin{align}
    &\alpha=1-\gamma, \qquad \delta=1-\beta,\qquad \beta,\gamma\in[0,1]\,.
    \label{stochasticcond}
\end{align}
We consider open boundary conditions, where at the boundaries we place two stochastic baths. We assume that the updates of boundary cells are local and Markovian.

As before, the propagator $\mathbb{U}$ is factored in two half-time steps
\begin{align}
    &\mathbb{U}=\mathbb{U}_\mathrm{o} \mathbb{U}_\mathrm{e},\label{bigUopen}\\&\mathbb{U}_\mathrm{e}=U_{123}U_{345}\dots U_{N-3,N-2,N-1}U_{N-1,N}^R\label{bigUeopen},\\
    &\mathbb{U}_\mathrm{o}=U_{12}^L U_{234}U_{456}\dots U_{N-4,N-3,N-2}U_{N-2,N-1,N},\label{bigUoopen}
\end{align}
where the local gate $U$ is given by \eqref{propagatorU}, with $f_{ij}$ given in \eqref{deformation}, along with the stochastic restriction on the parameters \eqref{stochasticcond}. As in the periodic case, $N$ is assumed to be even. The boundary gates $U_{12}^L$ and $U_{N-1,N}^R$ are $4\,\times \,4$ stochastic matrices acting on the first two or last two sites of the chain. In this work, we focus on the family of boundary baths 
referred to as \textit{conditional driving}~\cite{buvca2021rule}
\begin{align}
    &U^\rom{L}=\left(
\begin{array}{cccc}
 a & 0 & a & 0 \\
 0 & b & 0 & b \\
 1-a & 0 & 1-a & 0 \\
 0 & 1-b & 0 & 1-b \\
\end{array}
\right),
&&U^\rom{R}=\left(
\begin{array}{cccc}
 c & c & 0 & 0 \\
 1-c & 1-c & 0 & 0 \\
 0 & 0 & d & d \\
 0 & 0 & 1-d & 1-d \\
\end{array}
\right),
\label{conditionaldriving}
\end{align}
where $a, b, c, d\,\in\,(0,1)$ are the driving probabilities. The term conditional driving refers to the fact that the probability of changing site $1$ depends only on the state of site $2$ (similarly, for the right boundary, the probability of changing site $N$ depends only on the state of site $N-1$). For example, if site $2$ is empty (occupied), site $1$ will be empty with probability $a$ ($b$) or will contain a particle with probability $1-a$ ($1-b$). Another, distinct family of exactly solvable boundary conditions has been proposed~\cite{prosen2016integrability} and later generalized in \cite{inoue2018two}; however, we do not consider it in this work, although we expect that variants of all the results hold in that case as well.

\begin{figure}[H]
    \centering    \includegraphics[width=0.5\textwidth]{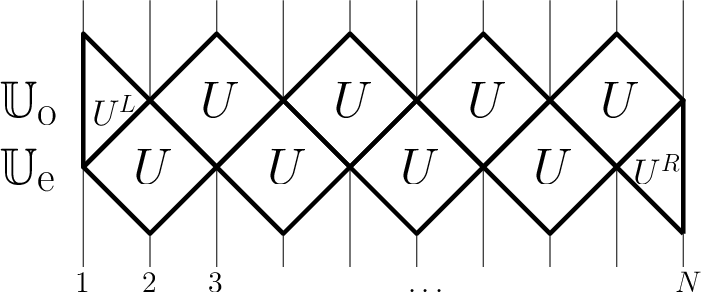}
    \caption{Graphical representation of the dynamics \eqref{bigUopen}. Time runs bottom-up.}
    \label{figurematrices3}
\end{figure}

In the context of the open boundary driving problem, we refer to integrability as the ability to find an expression for the non-equilibrium steady state (NESS) $\mathbf{p}$ in closed form via a matrix-product ansatz. The NESS is a fixed point of the dynamics,
\begin{align}
\mathbb{U}\,\mathbf{p} = \mathbf{p}.
\end{align}
We discuss the explicit construction of the NESS in Sec. \ref{boundarydrivness}. Additionally, we implemented empirical integrability diagnostics to the entire spectrum of $\mathbb U$ using the complex spacing ratio method~\cite{sa2020complex} and found clear signatures of integrability.
\section{Bulk Integrability}
\label{bulksection}
\subsection{Recap on Integrability}
\label{recaponint}
A unique definition of quantum integrability is still missing~\cite{caux2011remarks}. However, all the definitions share two common features:
\begin{itemize}
    \item There should exist an extensive (in $N$) set of local or quasi-local conserved charges $Q_i$ that commute with the propagator $\mathbb{U}$
    \begin{align}
        &[\mathbb{U},Q_i]=0,\,\,i=1,2,\dots
    \end{align}
    \item There should exist stable quasi-particle excitations that scatter elastically, with multiparticle scattering factorizing into two-particle scatterings\footnote{We note that the relation between classical field-theory integrability and the absence of particle production has recently been revisited in the literature, where subtleties may arise in certain classically integrable models, see \cite{BorsatoParticle,deLeeuwParticle}. This discussion does not affect the quantum/lattice setting considered in the present work.}.
\end{itemize}
\subsubsection{Yang-Baxter integrability.} In this work, we focus on models that are Yang-Baxter integrable, characterized by a Lax operator $\La$ satisfying the so-called RLL relation
\begin{align}
    &R_{b,a}(v,u)\mathcal{L}_{b,j}(v)\mathcal{L}_{a,j}(u)=\mathcal{L}_{a,j}(u)\mathcal{L}_{b,j}(v)R_{b,a}(v,u).
    \label{RLLNNmodel}
\end{align}
This is a matrix relation defined in the triple tensor product vector space $\text{End}(V\otimes V \otimes W)$, with $V\equiv \mathbb{C}^D$ and $W\equiv \mathbb{C}^d$, i.e., $D$- and $d$-dimensional complex vector spaces. $a,b$ and $j$ identify the spaces where the operators $\mathcal{L}$ and $R$ are acting non-trivially, $V$ is the auxiliary space and $W$ is the physical space, $u,v$ are known as spectral parameters and can take values in $\mathbb{C}$. The $R$-matrix is a solution of the Yang-Baxter equation
\begin{equation}
R_{12}(u_1,u_2)R_{13}(u_1,u_3)R_{23}(u_2,u_3)=R_{23}(u_2,u_3)R_{13}(u_1,u_3)R_{12}(u_1,u_2),
\label{ybennmodel}
\end{equation}
defined in $\text{End}(V\otimes V \otimes V)$.

By repeatedly using the RLL relation \eqref{RLLNNmodel}, the following commutativity relation can be explicitly proven
\begin{align}
    &[t(u),t(v)]=0, \label{ttcom}
\end{align}
where the transfer matrix $t(u)\in \text{End}\big(W^{\otimes N}\big)$ reads
\begin{align}
t(u)=\text{tr}_{a} \mathcal{L}_{aN}(u)\dots \mathcal{L}_{a2}(u)\mathcal{L}_{a1}(u),
\end{align}
with $\tr_{a}$ denoting the partial trace over the auxiliary space ${a}\equiv V$.

$t(u)$ can be interpreted as the generating function of conserved charges $Q_n$ of the model. By expanding the logarithm of $t(u)$ in a power series and defining the charges $Q_n$ as the coefficients of $u^{n-1}$, 
\begin{align}
Q_n = 
\partial^{n-1}_u \log t(u)|_{u=0},
\label{chargesQi}
\end{align}
Eq.~(\ref{ttcom}) implies the involutivity of the charges
\begin{align}
[Q_i,Q_j]=0,\quad i,j=2,3,\ldots
\end{align}
The charges $Q_i$ are translationally invariant sums of local operators with interaction range $i$, provided that the Lax operator satisfies the regularity condition $\mathcal{L}_{ij}(0)=P_{ij}$, where $P_{ij}$ is the transposition permutation (aka SWAP) operator.
Usually, one identifies $Q_2$ as an integrable Hamiltonian
\begin{align}
     &Q_2=\sum_{k=1}^N h_{k,k+1},\,\,\,\,\,\,\,\,\,\,h_{N,N+1}\equiv h_{N,1}.
\end{align}
The expressions of the higher conserved charges $Q_i$ can be either derived from the expression \eqref{chargesQi} or by using the boost operator\footnote{We will not discuss this second approach in detail here, as our goal is to provide a brief introduction to Yang-Baxter integrability for nearest-neighbour models and then extend the discussion to medium-range models, for which an explicit expression of the boost operator is still unknown. For further details, we refer the reader to \cite{Tetelman, HbbBoost} and more recently to \cite{deLeeuw:2020ahe,classificationybandboost}.} $\mathcal{B}[Q_2]$
\begin{align}
    &Q_{n+1}=[\mathcal{B}[Q_2],Q_n], \label{boostaction}\\
    &\mathcal{B}[Q_2]=\partial_u+\sum_{k=-\infty}^\infty k \, h_{k,k+1}   \label{booststandard},
\end{align}
which can, formally, only be defined on an infinite lattice ($N=\infty$), while its action  on conserved charges (\ref{boostaction}) is well (uniquely) defined for any finite periodic lattice. For example, the expression of $Q_3$ for a periodic spin chain is
\begin{align}
    &Q_3=\sum_{i=1}^{N}\left([h_{i,i+1},h_{i+1,i+2}]+\partial_u^2 \mathcal{L}_{i,i+1}(u)|_{u=0}\right).
    \label{Q3NNmodel}
\end{align}

Many studies have focused on integrable models where the $R$-matrix and the Lax operator act on two consecutive sites of the lattice, known as nearest-neighbour (n.n.) interactions. In \cite{sajat-medium}, this formalism was extended to models where the Lax operator acts on a finite number of consecutive sites greater than two. Gombor and Pozsgay refer to this class of translationally invariant models as \textit{medium-range models}, and we adopt this terminology in the present work. Here, we present only the expressions relevant to our discussion; for a more detailed explanation, we refer the reader to their original work.
\subsubsection{Medium range integrability.}
For medium range model, the RLL equation and the expression for the transfer matrix follow naturally from the n.n. case. Specifically, the RLL relation takes the following form
\begin{align}
    &R_{B,A}(v,u)\mathcal{L}_{B,j}(v)\mathcal{L}_{A,j}(u)=\mathcal{L}_{A,j}(u)\mathcal{L}_{B,j}(v)R_{B,A}(v,u),
    \label{RLLmediummodel}
\end{align}
 where now $A,B$ identify a pair of auxiliary spaces $V_A=V_B=V_{{a}_1}\otimes V_{{a}_2}\otimes \cdots \otimes V_{{a}_{q-1}}$ with ${q}$ the interaction range of the model\footnote{We define the interaction range of the model as the range of the first charge that generates the dynamics.}. We use capital letters to denote the tensor product of $(q-1)$ elementary spaces (i.e., $D=d^{q-1}$). The Yang-Baxter equation (YBE) retains the same form as in \eqref{ybennmodel}, with the substitutions $1 \to A$, $2 \to B$, $3 \to C$, to allow gluing of different spaces as for \eqref{RLLmediummodel}.

Given \eqref{RLLmediummodel}, the commutativity property $[t(u), t(v)] = 0$ holds when considering the transfer matrix
\begin{align}
    &t(u)=\tr_{A} \mathcal{L}_{A,N}(u)\mathcal{L}_{A,N-1}(u)\dots \mathcal{L}_{A,2}(u)\mathcal{L}_{A,1}(u).
    \label{transfermedium}
\end{align}
The conserved charges can also be constructed by taking the logarithmic derivative of \eqref{transfermedium}. The charges are local provided the regularity condition\footnote{For medium range model, the Lax operator has to reduce to a multiple-transposition permutation operator that has the role of swapping the auxiliary and the physical spaces.} holds
\begin{align}
    &\mathcal{L}_{a_1,a_2,\cdots,a_{q-1},j}(0)=P_{a_1,j}P_{a_2,j}\cdots P_{a_{q-1},j}.
\end{align}

In Sec. \ref{evidenceofintegrability}, point 3., we focus explicitly on the expressions of some of the conserved charges of the deformed RCA54. These expressions are general for a model with a Hamiltonian of range-3 interaction\footnote{Specifically, for the deformed RCA54, by gluing nearby sites, we bring the charge $Q_6$ that is invariant to shifts by two sites to a range-3 translational invariant charge.}. In particular, by formally taking the logarithmic derivative of the transfer matrix, we construct the first two conserved charges for this model. Furthermore, we also obtain the analytical expression of the third conserved charge. Computing the logarithmic derivative directly would be lengthy and tedious. As we will explain, we instead obtained it by performing multiple gluings of the spaces, which maps the model to a nearest-neighbor model in an enlarged Hilbert space.

\subsubsection{Integrable quantum circuits.}
Following the work \cite{vanicat2018integrable2}, one possible way to realize an integrable quantum circuit is to use the Trotterization procedure. This idea goes back to the light cone regularization of integrable QFTs, \cite{destri1987light}. 
The $R$-matrix can be used as a two-site quantum gate, and in certain cases this leads to discrete unitary time evolution with well defined integrability properties. The initial translationally invariant spin chains can be recovered in the continuous time limit of the quantum circuit models. A similar method can be used for medium range models, hence considering the Lax operator that satisfies the RLL relation \eqref{RLLmediummodel} as the quantum gate. Some models of this class, where the Lax operator is acting on $\mathbb{C}^2\otimes\mathbb{C}^2\otimes \mathbb{C}^2$, have been classified in \cite{sajat-medium} and include some of the reversible cellular automata classified by Bobenko et al.~\cite{bobenko1993two}.

However, the simplest deterministic model, RCA54, and its (quantum) deformation do not fit this framework: the dynamical evolution operator ${U}_{x-1,x,x+1}$ cannot be identified with a range-3 Lax operator acting on $\mathbb{C}^2 \otimes \mathbb{C}^2 \otimes \mathbb{C}^2$ that satisfies the RLL relation. A clear indication of this is that the first local (non diagonal) charge commuting with the dynamical evolution $\mathbb{U}$ of both the deformed and undeformed RCA54 models has interaction range 6. In \cite{gombor2024integrable}, the integrability of the undeformed RCA54 model was demonstrated using a Lax operator acting on six sites. According to the authors, the constructed transfer matrix does not directly generate the circuit evolution. We now turn to the deformed RCA54 model, for which the charge $Q_6$ is translational invariant with respect to two-site shifts for which the medium-range spin-chain approach cannot be applied in its original form. Nevertheless, we show that by performing 
a two-step gluing procedure between n.n. sites, the medium-range spin-chain formalism can still be utilized.

\subsection{Framework for proving integrability of the deformed RCA54 model}
\label{frameworkint}
We demonstrate how to embed the deformed RCA54 model into the Yang-Baxter integrability framework, where the local physical space is a qubit, $d=2$. In particular, we show that, for any choice of the deformation parameters, the first non-trivial conserved charge (with interaction range 6) is part of a hierarchy of commuting charges generated by higher-order logarithmic derivatives of the transfer matrix. This allows us to reformulate the integrability problem of the deformed RCA54 model as the following question:

\vspace{0.2cm}

\textit{Is there a transfer matrix that commutes with $\mathbb{U}$, the operator generating discrete-time dynamics of the deformed RCA54, such that its (higher-order) logarithmic derivatives yield the full tower of conserved charges of the model?}

\vspace{0.2cm}

We address this question in two ways. First, we establish a positive result within perturbation theory,
expressing the Lax operator up to the third order in the spectral
parameter $u$, for variable (symbolic, unspecified) deformation parameters. 
In this setup, we prove that the charge $Q_6$ belongs to an infinite tower of commuting conserved charges. 
However, the fact that the charges commute with the deformed RCA54 propagator $\mathbb U$ can only be proven perturbatively up to the third order (i.e., for the first three local charges).

Second, we prove integrability by constructing the full non-perturbative, spectral-parameter-dependent Lax operator for a fixed generic choice of (for example, rational) deformation parameters. 
Our algorithm can be used to obtain the Lax operator for different specific numerical choices. 
However, a full analytical proof is not available at present. 
This approach allows us to demonstrate that the charges commute and are conserved.

In the following, we elaborate on both the perturbative and non-perturbative (exact) methods.
 In Secs. \ref{perturbativesketch} and \ref{analyticsketch}, we outline the key steps of our approach; for a more detailed technical exposition, we refer the reader to Secs. \ref{evidenceofintegrability1} and \ref{technicaldetailsLax}.

\subsubsection{Perturbative proof up to third order (for variable deformation parameters)}
\label{perturbativesketch}
We implement the following steps:
\begin{itemize}
    \item[1.] We find the conserved charge of interaction range 6, $Q_6$ such that $[\mathbb{U},Q_6]=0$. $Q_6$ is translationally invariant under a shift by two lattice sites.
    \item[2.] We glue nearest-neighbour sites of the original model, hence we consider $Q_6$ as a charge of interaction range 3 (in the glued picture). We refer to it as $\mathcal{Q}_3$.
    \begin{itemize}
        \item [2.1] We strictly follow the medium-range spin-chain formalism of \cite{sajat-medium} and prove that the charge 
        $\mathcal{Q}_{5}$ (corresponding to $Q_{10}$ in the original picture), constructed by taking an additional formal derivative of the transfer matrix, commutes with it.
        \item [2.2] We check that\footnote{For brevity, in all pictures, whether glued or unglued, we denote the time-evolution operator of the deformed rule-54 model by $\mathbb{U}$. In \ref{algebraicprooftuU}, we explicitly shows the expression for $\mathbb{U}$ in the glued picture.} $[\mathbb{U},\mathcal{Q}_5]=0$. 
    \end{itemize}
\end{itemize}
At this point, we can rely on the recent proof of the star-triangle conjecture for nearest-neighbor models by A. Hokkyo \cite{hokkyo2025integrability}. This conjecture, originally proposed in Ref.~\cite{Idzumi:1994kx}, states that if a model possesses two translationally invariant conserved charges, one of which can be obtained from the other via the boost operator, then the model is Yang--Baxter integrable. An important caveat is that the proof applies provided that there exist no additional nearest-neighbor quantities commuting with the Hamiltonian. The deformed RCA54 model may appear to lie outside this class, since its first conserved charge has range 6 (or range 3 after gluing nearby sites). However,

\begin{itemize}
    \item [3.] We take the glued chain and glue it once more resulting in a $2^4=16$-dimensional local Hilbert space.  In this way, we can treat $\mathcal{Q}_3$ as a range-2 charge $\Theta_2$ and similarly $\mathcal{Q}_5$ as a range-3 charge $\Theta_3$. At this point, the result of \cite{hokkyo2025integrability} guarantees the existence of an infinite number of charges\footnote{We remark that, in this double-glued picture, the charge $\Theta_2$ is the only one of range 2.}. Since we are interested in checking the commutator with $\mathbb{U}$, we again use the boost operator in its standard form \eqref{booststandard} to construct the next charge, $\Theta_4$ (corresponding to $Q_{14}$).
    \begin{itemize}
        \item [3.1] We explicitly verify that $[\mathbb{U},\Theta_4]=0$.
    \end{itemize}
    \item[4.] We show how to use the three conserved charges $Q_6, Q_{10}$ and $Q_{14}$ to obtain a perturbative expansion of the Lax operator up to third order in the spectral parameter.
\end{itemize}
To summarize, we prove that, given
\begin{align}
    &[\mathbb{U},Q_6]=0,
\end{align}
the following also holds
\begin{align}
    &[Q_6,Q_{10}]=[Q_6,Q_{14}]=[Q_{10},Q_{14}]=0,
    \label{Q6Q10Q14}
    \end{align}
where $Q_{10}$ and $Q_{14}$ are constructed by formally taking the logarithmic derivative of the transfer matrix \eqref{transfermedium}, and hence depend on $Q_6$, together with additional range-6 terms to be determined\footnote{This statement is explained in Sec. \ref{evidenceofintegrability}, point 4.}. 

Equivalently, in the double glued picture, these charges correspond to charges with lower range interaction in a higher dimensional Hilbert space
\begin{align}
    &[\Theta_2,\Theta_3]=[\Theta_2,\Theta_4]=[\Theta_3,\Theta_4]=0,
\end{align}
where $\Theta_i=\sum_{k}\theta^{(i)}_{[k,k+i-1]}$, $[k,k+i-1]\equiv k,k+1,\dots, k+i-1$.

In this higher dimensional Hilbert space, one can define a boost operator analogous to \eqref{booststandard}, where $h_{k,k+1}$ is now replaced by $\theta^{(2)}_{k,k+1}$
\begin{align}
    &\Theta_3=\big[\mathcal{B}[\Theta_2],\Theta_2],
\end{align}
hence, the proof of the star-triangle conjecture of Ref.~\cite{hokkyo2025integrability}, implies that
\begin{align}
    &[\Theta_i,\Theta_j]=0, \,\,\forall \, i,j=2,3,\dots 
\end{align}
Furthermore, we checked that
\begin{align}
    &[\mathbb{U},Q_{10}]=[\mathbb{U},Q_{14}]=0\,.
\end{align}

The perturbative expansion of the Lax operator reads
\begin{equation}
    \check{\mathcal{L}}_{i,i+1,i+2}(u) = P_{i+1,i+2} P_{i,i+2} {\mathcal{L}}_{i,i+1,i+2}(u)
    = 1 + u \mathrm{h}_{i} + \frac{u^2}{2}(\tilde{\mathrm{h}}_{i} + \mathrm{h}_{i}^2) 
    + \frac{u^3}{3!}(\mathrm{h}_{i}^3 + \mathrm{h}_{i}\tilde{\mathrm{h}}_{i} + 2\tilde{\mathrm{h}}_{i}\mathrm{h}_{i} - 2\tilde{\tilde{\mathrm{h}}}_{i}) + O(u^4),
    \label{ansatzL1}
\end{equation}
For brevity, we denote $\mathrm{h}_i \equiv \mathrm{h}_{i,i+1,i+2}$, and similarly for $\tilde{\mathrm{h}}_i$ and $\tilde{\tilde{\mathrm{h}}}_i$. 

The range-3 glued operator $\rom{h}_i$ (range 6 in the original picture, i.e., a $64 \times 64$ matrix) represents the density of the charge $Q_6$. 
The operators $\tilde{\rom{h}}_{i}$ and $\tilde{\tilde{\rom{h}}}_{i}$ are also range-6 operators in the original picture, determined by the commutation relations \eqref{Q6Q10Q14}. 
Their explicit forms are provided in \eqref{Q5standardspace} and \eqref{Q7standardspace}.

This implies that
\begin{align}
    &[\log t(u),\mathbb{U}]=O(u^4)\,,
\end{align}
where $t(u)$ is the transfer matrix defined in \eqref{transfermedium} and explicitly written in \eqref{transferrange3} with auxiliary space $\mathbb{C}^4\otimes\mathbb{C}^4$.

In Sec. \ref{evidenceofintegrability}, we elaborate on each of these steps, providing additional theoretical background where necessary. 
A graphical illustration of the two gluings is shown in the figure below:

\vspace{0.5cm}

\noindent
\begin{tcolorbox}[colback=white,colframe=black,width=\textwidth,boxrule=0.8pt,halign=center]
\begin{minipage}[t]{0.48\textwidth}
\vspace{0pt}
\textbf{Point 1:}

original model
$$Q_6=\sum_{i=1}^{N/2}q_{[2i-1,2i+4]}$$

{\begin{tikzpicture}[baseline={(0,0)}]
    \fill (0,0) circle (0.07); 
    \node[right=2mm] at (0,0) {$= \mathbb{C}^2$};
\end{tikzpicture}
}

\end{minipage}%
\hfill
\begin{minipage}[t]{0.48\textwidth}
\vspace{10pt}
\centering
\resizebox{\linewidth}{!}{
\begin{tikzpicture}[scale=1]

\def\cx{4.5}        
\def\cy{0}          
\def\radiusx{4.5}   
\def\radiusy{1.25}  

\draw[dashed] (\cx-\radiusx,\cy) 
    arc[start angle=180,end angle=0,x radius=\radiusx,y radius=\radiusy];

\draw[solid] (\cx-\radiusx,\cy) 
    arc[start angle=180,end angle=360,x radius=\radiusx,y radius=\radiusy];

\foreach \i in {0,1,...,7} {
    \pgfmathsetmacro{\theta}{180 + (\i+0.5)*180/8} 
    \pgfmathsetmacro{\x}{\cx + \radiusx*cos(\theta)} 
    \pgfmathsetmacro{\y}{\cy + \radiusy*sin(\theta)} 
    \fill (\x,\y) circle (0.1); 
}

\foreach \i/\label in {0/N,1/1,2/2,3/3,4/4,5/5,6/6,7/{...}} {
    \pgfmathsetmacro{\theta}{180 + (\i+0.5)*180/8}
    \pgfmathsetmacro{\x}{\cx + \radiusx*cos(\theta)} 
    \pgfmathsetmacro{\y}{\cy + \radiusy*sin(\theta)}
    \node at (\x, \y-0.3) {\label};
}

\pgfmathsetmacro{\thetaStart}{180 + 1.5*180/8}
\pgfmathsetmacro{\thetaEnd}{180 + 6.5*180/8}
\pgfmathsetmacro{\xstart}{\cx + \radiusx*cos(\thetaStart)}
\pgfmathsetmacro{\xend}{\cx + \radiusx*cos(\thetaEnd)}
\pgfmathsetmacro{\ystart}{\cy + \radiusy*sin(180 + 4.0*180/8) - 0.5} 
\draw[decorate,decoration={brace,mirror,amplitude=5pt}] 
    (\xstart, \ystart) -- (\xend, \ystart) node[midway,below=6pt] {$q_{[1,6]}$};

\end{tikzpicture}%
}
\end{minipage}
\end{tcolorbox}

\noindent
\begin{tcolorbox}[colback=white,colframe=black,width=\textwidth,boxrule=0.8pt,halign=center]
\begin{minipage}[t]{0.48\textwidth} 
\vspace{0pt}
\textbf{Point 2:}

first gluing
$$\mathcal{Q}_3=\sum_{j=1}^{N/2}\rom{q}_{[j,j+2]}=\sum_{j=1}^{N/2}\rom{h}_{j}$$

\begin{tikzpicture}[baseline={(0,0)}]
    \draw[fill=white,line width=0.5pt] (0,0) circle (0.15);

    \node[right=4mm] at (0,0) {$=$};

    \draw[fill=black,line width=0.8pt] (1.2,0) circle (0.065);

    \node[right=4mm] at (1.2,0) {$\otimes$};

    \draw[fill=black,line width=0.8pt] (2.4,0) circle (0.065);
    \node[right=4mm] at (2.4,0) {$=\,\,\mathbb{C}^4$};
\end{tikzpicture}

\end{minipage}%
\hfill
\begin{minipage}[t]{0.48\textwidth}
\vspace{10pt}
\centering
\resizebox{\linewidth}{!}{%
\begin{tikzpicture}[scale=1]

\def\cx{4.5}        
\def\cy{0}          
\def\radiusx{4.5}   
\def\radiusy{1.25}  

\draw[dashed] (\cx-\radiusx,\cy) 
    arc[start angle=180,end angle=0,x radius=\radiusx,y radius=\radiusy];

\draw[solid] (\cx-\radiusx,\cy) 
    arc[start angle=180,end angle=360,x radius=\radiusx,y radius=\radiusy];

\foreach \i/\label in {0/{\text{N/2}},1/{1},2/{2},3/{3},4/{...},5/{...}} {
    \pgfmathsetmacro{\theta}{180 + (\i+0.5)*180/6}
    \pgfmathsetmacro{\x}{\cx + \radiusx*cos(\theta)} 
    \pgfmathsetmacro{\y}{\cy + \radiusy*sin(\theta)}

    \draw[fill=white,line width=0.8pt] (\x,\y) circle (0.2); 

    \node at (\x, \y-0.5) {\fontsize{12}{14}\selectfont \label};
}

\pgfmathsetmacro{\thetaStart}{180 + 1.5*180/6}
\pgfmathsetmacro{\thetaEnd}{180 + 3.5*180/6}
\pgfmathsetmacro{\xstart}{\cx + \radiusx*cos(\thetaStart)}
\pgfmathsetmacro{\xend}{\cx + \radiusx*cos(\thetaEnd)}
\pgfmathsetmacro{\ystart}{\cy - \radiusy - 0.7}
\draw[decorate,decoration={brace,mirror,amplitude=5pt}] 
    (\xstart, \ystart) -- (\xend, \ystart) node[midway,below=6pt] {$\rom{q}_{[1,3]}$};

\end{tikzpicture}
}
\end{minipage}
\end{tcolorbox}

\begin{tcolorbox}[colback=white,colframe=black,width=\textwidth,boxrule=0.8pt,halign=center]
\begin{minipage}[t]{0.48\textwidth}
\vspace{0pt}
\textbf{Point 3:}

second gluing
$$\Theta_2=\sum_{j=1}^{N/4}\eta_{j,j+1}$$

\begin{tikzpicture}[baseline={(0,0)}]
    \draw[fill=white,line width=0.71pt] (0,0) circle (0.2);

    \node[right=4mm] at (0,0) {$=$};

    \draw[fill=white,line width=0.65pt] (1.2,0) circle (0.13);

    \node[right=4mm] at (1.2,0) {$\otimes$};

    \draw[fill=white,line width=0.65pt] (2.4,0) circle (0.13);
    \node[right=4mm] at (2.4,0) {$=\,\,\mathbb{C}^{16}$};
\end{tikzpicture}

\end{minipage}%
\hfill
\begin{minipage}[t]{0.48\textwidth}
\vspace{10pt}
\centering
\resizebox{\linewidth}{!}{%
\begin{tikzpicture}[scale=1]

\def\cx{4.5}        
\def\cy{0}          
\def\radiusx{4.5}   
\def\radiusy{1.25}  

\draw[dashed] (\cx-\radiusx,\cy) 
    arc[start angle=180,end angle=0,x radius=\radiusx,y radius=\radiusy];

\draw[solid] (\cx-\radiusx,\cy) 
    arc[start angle=180,end angle=360,x radius=\radiusx,y radius=\radiusy];

\foreach \i/\label in {0/{\text{N/4}},1/{1},2/{2},3/{3}} {
    \pgfmathsetmacro{\theta}{180 + (\i+0.5)*180/4}
    \pgfmathsetmacro{\x}{\cx + \radiusx*cos(\theta)} 
    \pgfmathsetmacro{\y}{\cy + \radiusy*sin(\theta)}

    \draw[fill=white,line width=0.8pt] (\x,\y) circle (0.3); 

    \node at (\x, \y-0.6) {\fontsize{12}{14}\selectfont \label};
}

\pgfmathsetmacro{\thetaStart}{180 + 1.5*180/4}
\pgfmathsetmacro{\thetaEnd}{180 + 2.5*180/4}
\pgfmathsetmacro{\xstart}{\cx + \radiusx*cos(\thetaStart)}
\pgfmathsetmacro{\xend}{\cx + \radiusx*cos(\thetaEnd)}
\pgfmathsetmacro{\ystart}{\cy - \radiusy - 0.9}
\draw[decorate,decoration={brace,mirror,amplitude=5pt}] 
    (\xstart, \ystart) -- (\xend, \ystart) node[midway,below=6pt] {$\eta_{1,2}=\rom{h}_{1}+\rom{h}_2$};

\end{tikzpicture}%
}
\end{minipage}
\end{tcolorbox}

\subsubsection{Non-perturbative proof (for a fixed choice of the deformation parameters)}
\label{analyticsketch}

Here we sketch the algorithm used to obtain the exact form of the Lax operator that generates an infinite series of commuting charges $Q_6,\,Q_{10},\,Q_{14},\,\dots$. 
We provide a graphical proof that the transfer matrices built from this Lax operator commute for different values of the spectral parameter (see \ref{prooftutv}), and that the transfer matrix commutes with the discrete-time propagator, i.e. the corresponding local charges are conserved (see \ref{prooftuU}). 
In Sec. \ref{technicaldetailsLax}, we present a more detailed description of the algorithm, together with an algebraic proof of commutativity and charge conservation using explicit index notation.

For simplicity, we made an explicit numerical choice of the parameters
\begin{align}
    &\alpha =\frac{1}{7};&&\beta =\frac{1}{2};&&\gamma =\frac{1}{8};&&\delta =\frac{3}{11}.
    \label{numericalchoice}
\end{align}
We deliberately chose a deformation that makes the gate $U$ neither unitary nor stochastic, as we believe that the origin of integrability is purely algebraic. 
We remark that our algorithm can also be used (and has been tested) for other rational numerical choices of the deformation parameters. Nevertheless, we expect the Mathematica code to work for exact irrational parameter values, such as algebraic numbers, although the computations would be considerably slower.

\begin{figure}[H]
    \centering    \includegraphics[width=0.5\textwidth]{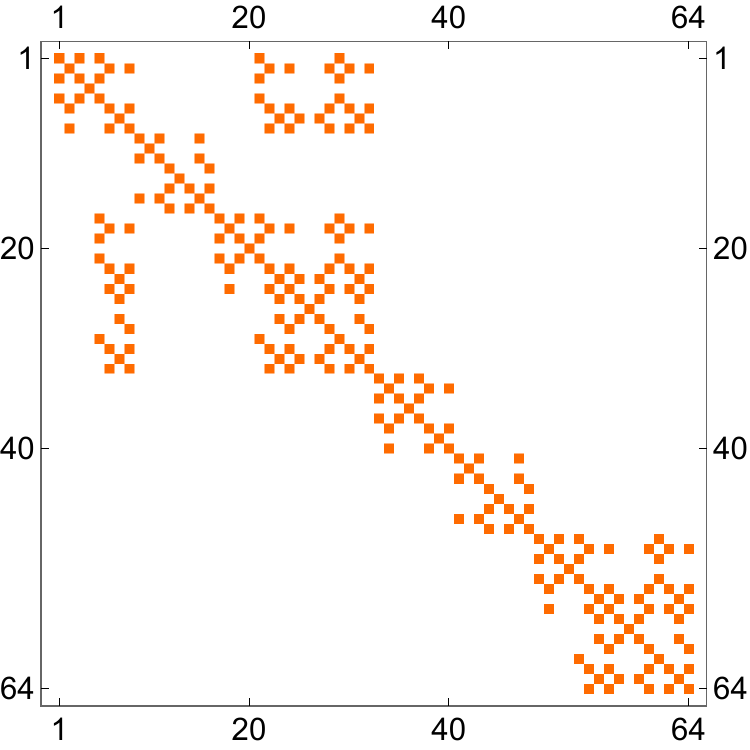}
    \vspace{-0.4cm}
    \caption{Non-zero entries of the matrix representation of the Lax operator $\check{\mathcal{L}}(u)$ are shown. From its structure, it is clear that the matrix is diagonal with respect to the first and last (sixth) qubits, hence it has the structure of a face tensor.}
    \label{nonzeroLax}
\end{figure}

Finding the Lax operator (a $2^6\times2^6$ matrix), which is the building block of the transfer matrix \eqref{transfermedium} and satisfies $[t(u),Q_6]=0$, is computationally very demanding. 
To simplify this task, we proceed in the following steps:
\begin{itemize}
    \item [1.] First, starting from the perturbative expression of the Lax operator \eqref{ansatzL1}, we construct an ansatz for the non-zero coefficients appearing in the higher-order terms
    
\begin{align}
    \check{\mathcal{L}}(u)=1+\sum_{n\ge 1} \frac{u^n}{n!} \check{\mathcal{L}}^{(n)}(0), 
\end{align}
    with $\check{\mathcal{L}}^{(n)}(0)$, $n=1,2,3$ given in \eqref{ansatzL1} and $n>3$ to be determined.
    
    To achieve this, it is important to observe that at orders $u^2$ and $u^3$ the largest number of non-zero matrix entries arises from the matrix powers $\rom{h}_i^2$ and $\rom{h}_i^3$. This suggests that all terms of order $\sim u^n$ contain non-zero entries originating from $\rom{h}_i^n$. Moreover, we note that considering terms up to $\rom{h}_i^4$ is sufficient to capture all non-zero entries in $\rom{h}_i^n$ for $n>4$. 
   We therefore consider an ansatz in which the corresponding non-zero entries of the Lax matrix are replaced by arbitrary functions $f_k(u)$, with $k=1,\dots,\text{\# non-zero entries}$.

    \item[2.] Order by order in $u$, we solve the commutativity condition
    \begin{align}
        [t(u),Q_6]=0.
    \end{align}
    In this way, we obtain $\check{\mathcal{L}}^{(n)}(0)$ for $n\le 35$.
    \item[3.] At this point, we would like to resum this finite but high-order perturbative expansion and derive analytic expressions for all entries of the Lax matrix. We observe that some of the entries are polynomials of degree 4 in $u$, and therefore do not receive corrections at higher orders in $u$. For the remaining entries, we employ different strategies to obtain the complete Lax matrix:
    \begin{itemize}
        \item[a.] We can resum some of the entries exactly as ratios of two low order polynomials using the Padé approximation;
        \item[b.] We consider the ratios between all pairs of entries in the perturbative expansions and resum some of them again as rational functions using the Padé approximant;
        \item[c.] For small $N$ ($N=2,4$), we calculate two transfer matrices, one by using the perturbative expansion of the Lax operator (up to order $O(u^{36})$) and the other by using the ansatz\footnote{In this ansatz, we explicitly substituted the entries that were polynomials of maximum degree 4.} in terms of the unspecified functions $f_k(u)$. In this way, we fix some of the remaining non-zero entries. 
     \item[d.] The remaining entries are obtained by solving $[t(u),Q_6]=0$.
    \end{itemize}
 By combining points a, b, c, and d, we construct the full Lax operator, which is provided in the attached Mathematica notebook. 
For completeness, we show in Fig. \ref{nonzeroLax} all the non-zero matrix elements of the Lax operator.

 \item[4.]
Finally, we check for $N=8$ that $[t(u),\mathbb{U}]=0$. We emphasize that this condition was not imposed; it follows naturally. 
In Sec. \ref{technicaldetailsLax}, we explain why verifying $N=8$ is sufficient to ensure that the result holds for any $N$.
 \end{itemize}
 
    \vspace{-0.6cm}

\subsubsection{Proof that $[t(u),t(v)]=0$}

\label{prooftutv}

\begin{figure}[H]
    \raggedright
\includegraphics[width=0.7\textwidth]{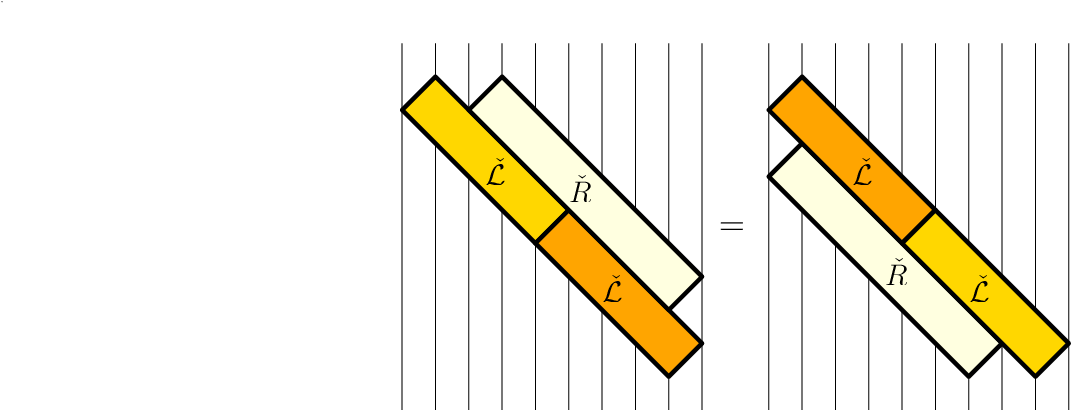}

\vspace{-0.3cm}

    \caption{Graphical representation of the equation \eqref{RLLcheck}. The orange and golden boxes represent range 6 operators $\check{\mathcal{L}}(u)$ and  $\check{\mathcal{L}}(v)$, respectively, acting in the vertical direction (over qubits as indicated by thin vertical lines). $\check{R}$ denotes the range 8 2-parameter operator $\check{R}(v,u)$.
    Note that all operators are diagonal in the first and the last qubit they act upon. In other words, if a corner of the tensor-box touches the thin vertical line, the corresponding qubit participates only as a control, and its state cannot change.
    }
\label{RLLpicture}
\end{figure}

This proof is standard and can be formulated as the existence of an intertwiner operator $\check{R}_{B,A}(v,u)$ that satisfies \eqref{RLLmediummodel} or equivalently
\begin{align}
    \check{R}_{[3-10]}(v,u)\check{\mathcal{L}}_{[1-6]}(v)\check{\mathcal{L}}_{[5-10]}(u)=\check{\mathcal{L}}_{[1-6]}(u)\check{\mathcal{L}}_{[5-10]}(v)\check{R}_{[1-8]}(v,u),\label{RLLcheck}
\end{align}
where the sites now refer to the original (unglued) local Hilbert spaces $\mathbb{C}^2$. Note that all operators ($\check{\mathcal{L}}$ and $\check{R}$) are diagonal in the first and the last qubit they act upon. Eq. \eqref{RLLcheck} is represented graphically in Fig. \ref{RLLpicture}, using, what may be interpreted as, a face version of tensor network notation\footnote{Face tensors are represented as rectangular plaquettes embedded within a regular lattice of vertical lines, where each line represents a local qubit.
Each rectangular plaquette operates as a matrix on a state of incoming lines, say coming from the bottom, yielding a state encoded in the outgoing lines at the top. Note that on the left or right side-corner the qubits act as controls as their states cannot be changed. Such a face tensor network notation is thus an efficient way to manipulate general left-control-right-control-gates. IRF formulation of the propagator of deformed RCA54 (Fig.~\ref{figurematrices2}) is a simple example of face tensor network. A related general formalism is known as shaded calculus~\cite{Vicary1,Vicary2}.}.

\vspace{-0.5cm}

The dimension of the $R$-matrix is $2^8\times 2^8$, and to demonstrate its existence and invertibility, we explicitly computed it by choosing specific numerical values for the
spectral parameters: $u=1/6$, $v=2/5$.

\begin{figure}[H]
    \centering
\includegraphics[width=0.6\textwidth]{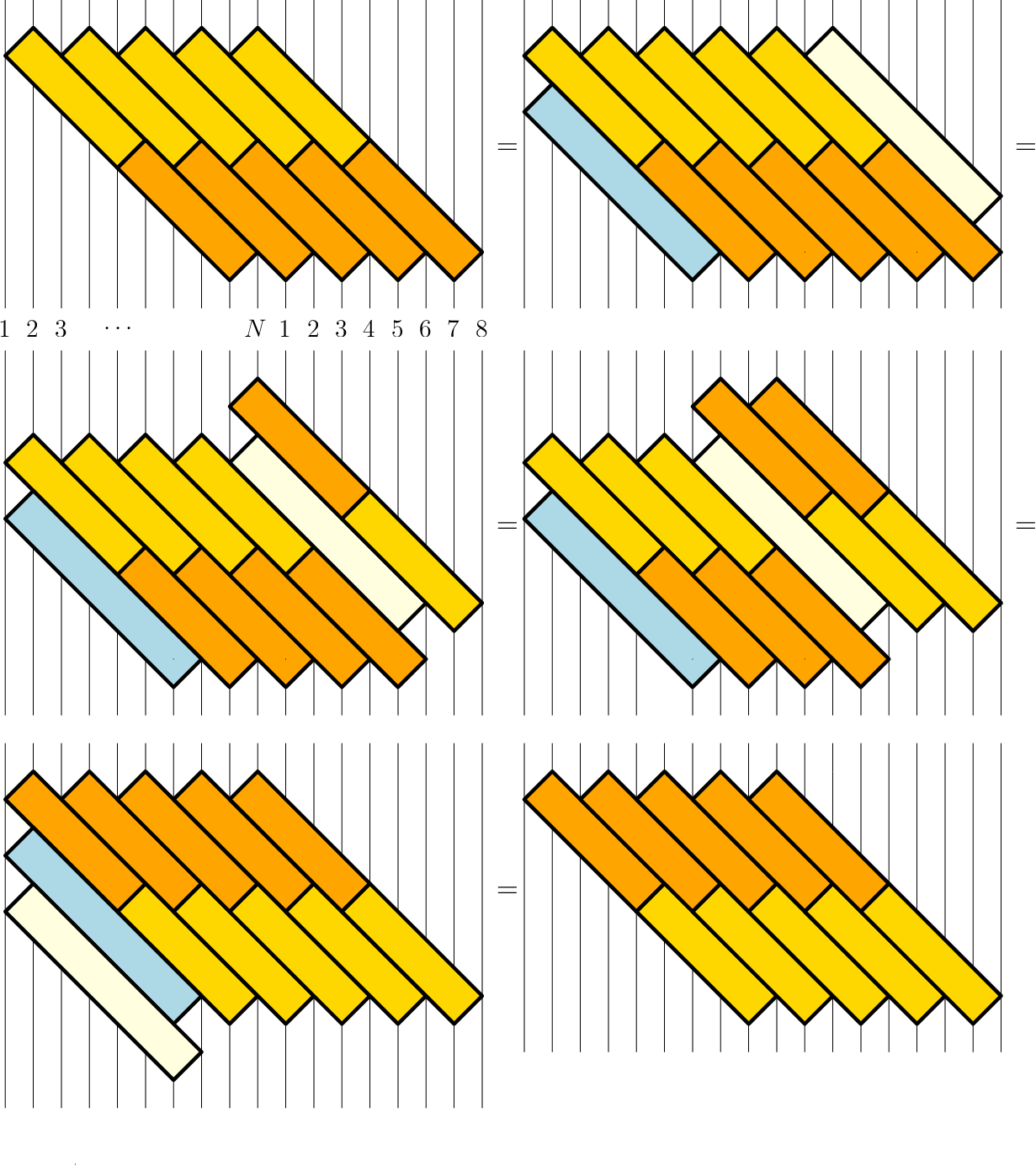}

\vspace{-0.8cm}

    \caption{Graphical proof that $[t(u),t(v)]=0$. The orange box represents $\check{\mathcal{L}}(u)$, the gold box represents $\check{\mathcal{L}}(v)$, the light yellow box represents $\check{R}(v,u)$, and the light blue box represents $\check{R}(v,u)^{-1}$. For simplicity, we present the proof for $N=10$, as the generalization to arbitrary $N$ is immediate. We remark that we consider a system with periodic boundary conditions. The algebraic proof can be found in Sec. \ref{algebraicprooftutv}.}
    \label{Rllproof}
\end{figure}

By repeatedly applying this relation, as illustrated in Fig. \ref{Rllproof}, we can prove graphically that $[t(u),t(v)]=0$.
The advantage of this proof, compared to those in Sec. \ref{perturbativesketch}, is that we do not need to perform additional gluing of spaces here. However, since we are at this point unable to provide a general variable parameter solution to Eq.~(\ref{RLLcheck}), such graphical proofs only apply to specific choices of deformation and spectral parameters.

\subsubsection{Proof that $[t(u),\mathbb{U}]=0$}
\label{prooftuU}
\begin{figure}[H]
    \centering    \includegraphics[width=0.55\textwidth]{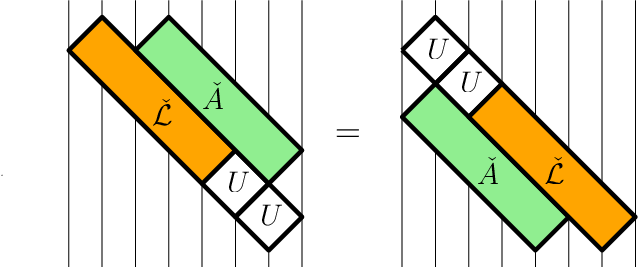}
    \caption{Graphical face-tensor representation of the equation \eqref{RLLUUa}.}
    \label{RLLUUfig}
\end{figure}

This proof is one of the main results of this work. The analogue of the RLL relation \eqref{RLLmediummodel} which is required to establish the commutation relation $[t(u),\mathbb{U}]=0$ is given by
\begin{equation}
    \check{A}_{[3-8]}\check{\mathcal{L}}_{[1-6]}U_{5,6,7}U_{6,7,8}=U_{1,2,3}U_{2,3,4}\check{\mathcal{L}}_{[3-8]}\check{A}_{[1-6]}, \label{RLLUUa}
\end{equation}
which is defined over $8$ qubit Hilbert space $(\mathbb{C}^2)^{\otimes 8}$ and is  represented graphically in face-tensor notation in Fig. \ref{RLLUUfig}.

\begin{figure}[H]
    \centering    \includegraphics[width=0.7\textwidth]{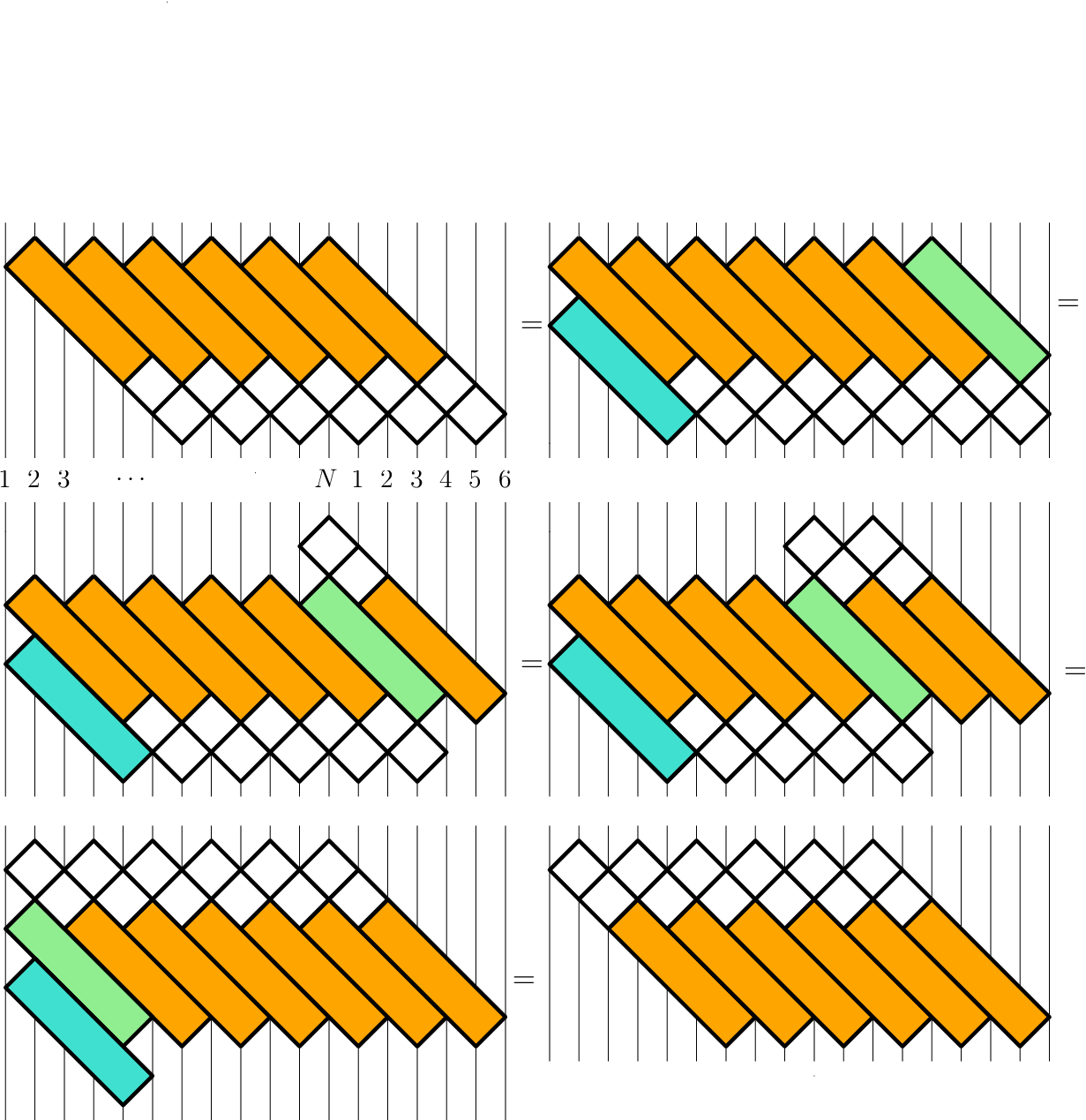}
    \caption{Graphical proof of identity $[t(u),\mathbb{U}]=0$. The orange boxes are $\check{\mathcal{L}}$, the white squares are $U$, the green box is $\check{A}$ and the turquoise one is $\check{A}^{-1}$. As it is clear from the diagrams, all the operators $\check{\mathcal{L}}$, $U$, $\check{A}$ and $\check{A}^{-1}$ are diagonal in the first and the last qubit. For simplicity, we present the proof for $N=12$, as the generalization to arbitrary $N$ is immediate. We remark that we consider a system with periodic boundary conditions. The algebraic proof can be found in Sec. \ref{algebraicprooftuU}.}
    \label{tuproof}
\end{figure}

We solve the linear equation \eqref{RLLUUa} for $\check{A}$ at two values of the spectral parameter $u$, $u=1/6$ and $u=2/5$. In both cases, we find that $\check{A}$ exists and is invertible. By repeatedly applying the graphical relation shown in Fig. \ref{RLLUUfig}, we then demonstrate graphically in Fig. \ref{tuproof} that \eqref{RLLUUa} is sufficient to imply $[t(u),\mathbb{U}]=0$.

\vspace{-2cm}

\subsection{Technical details of the steps of Sec. \ref{frameworkint}}
\label{evidenceofintegrability}
The following two sections present all the technical details supplementing Secs. \ref{perturbativesketch} and \ref{analyticsketch}.

\subsubsection{Perturbative proof of integrability of \ref{perturbativesketch}: step by step}

\label{evidenceofintegrability1}

\subsubsection*{1. Conserved charge $Q_6$.}
With the help of a linear equation solver in a computer algebra system (Mathematica), we obtain that the lowest range conserved charge generating the dynamics of deformed RCA54 model has interaction range 6,
\begin{align}
    &[Q_6,\mathbb{U}]=0,
    &&Q_6=\sum_{i=1}^{N/2} q_{[2i-1,2i+4]},
    \label{fullchargeQ6}
\end{align}
where $q_{[i,j]}$ denotes the density of the charge $Q_{j-i+1}$, acting nontrivially on sites $i,\dots,j$. The interval $[i,j]$ specifies the range of lattice sites  on which the density $q$ has nontrivial support. We note that, due to the even–odd layered structure of the time-evolution operator $\mathbb{U}$, the charge $Q_6$ is translationally invariant under shifts by two lattice sites.

Due to the structure of the local propagator $U$ (being diagonal on the first and the last qubit), the densities $q_{[2i-1,2i+4]}$ satisfy
\begin{align}
    &[q_{[1,6]},q_{[6,11]}]=0.
\end{align}
We provide the full expression of $q_{[2i-1,2i+4]}$ in \cite{mathematicanotebook} (\texttt{BulkIntegrability.nb}). Given \eqref{fullchargeQ6}, the expression for the density $q_{[1,6]}$ is not unique: it can be modified by an arbitrary term given by the difference between an operator and its two-site–shifted version. 
To fix this gauge freedom, we choose a representative of $q_{[1,6]}$ by retaining only those terms $P$ in the general Pauli operator expansion $q_{[1,6]}=\sum_P c_P P$ that do not begin with two identity operators on the leftmost sites, i.e. $\tr_{12} P = 0$. We refer to such an operator density as {\em left-aligned}.
For consistency, we verify that in the non-deformed limit, our range 6 operator (viewed as a translationally invariant charge at each site) reduces to the conserved charge density of \cite{gombor2024integrable}, upon setting \(\Delta = 1\) and applying an appropriate renormalization together with a shift by a term proportional to the identity. We note that this 
range 6 Hamiltonian was previously constructed in the context of coordinate Bethe ansatz for Rule 54 in Ref.~\cite{Friedman19}.

Furthermore, by direct computation one can obtain another charge, $Q_7$, with a nontrivial range 7 density that commutes with $\mathbb{U}$ (and also with $Q_6$). The corresponding density has the form $1_2\otimes \tilde{q}_6$, where $1_2$ denotes the identity on $\mathbb{C}^2$ and $\tilde{q}_6$ is obtained from $q_6$ by reflecting the spatial indices $i\to 7-i$, in the Pauli operator expansion (possibly followed by left-aligning all terms via two-site shifts). This spatial-reflection mapping extends to higher local charges, which therefore always appear in spatially symmetric doublets. Therefore, we will not consider in the following the dual family generated by $Q_7$.

\subsubsection*{2. Gluing of pairs of neighbouring local spaces, yielding $Q_6\,\to\,\mathcal{Q}_3$.}
We now show how to relate the conserved charge $Q_6$ to a Hamiltonian of a homogeneous spin chain and in step 2.1, using the formalism of medium range integrable spin chains~\cite{sajat-medium}, we take the formal logarithmic derivative of the transfer matrix to obtain the next conserved charge. Iterating this procedure generates the full tower of conserved charges, thereby establishing the integrability of the spin-chain model.

Since $Q_6$ is invariant under a two site shift, it is natural to interpret it as a range-3 operator acting in $\mathcal{H}\otimes\mathcal{H}\otimes \mathcal{H}$, where $\mathcal{H}=\mathbb{C}^2\otimes \mathbb{C}^2=\mathbb{C}^4$. The advantage of this gluing procedure is that it restores one-site translational invariance. The drawback, however, is that the local Hilbert space dimension is squared (from $2\to 4$).

From this point on, we identify $\mathcal{Q}_3=Q_6$ and adopt the notation $\mathcal{Q}$ and $\mathrm{q}$ for conserved charges and their corresponding densities with respect to the enlarged local Hilbert space $\mathbb C^4$, respectively. For $\mathcal{Q}_3$, we use the special notation $\mathrm{h}$ for its density, and introduce the shorthand  $\mathrm{h}_i=\mathrm{h}_{i,i+1,i+2}$.

\subsubsection*{2.1 Build $\mathcal{Q}_5$ from $\mathcal{Q}_3$.}

We proceed to show that there exists a range 5 conserved charge (range 10 in the original formulation) $\mathcal{Q}_5$ that commutes with $\mathcal{Q}_3$. Before presenting the result of this computation, we provide some background on the objects involved. Since the conserved range $\mathcal{Q}_3$ has interaction range 3, it is natural to assume that it originates from a Lax operator acting on three sites. The object responsible for generating the conserved charges is the transfer matrix\footnote{From \eqref{transfermedium} one identifies the auxiliary space $A=\{a,b\}$.}
\begin{equation}
    t(u)=\text{tr}_{a,b} \mathcal{L}_{ab,\frac{N}{2}}(u)\mathcal{L}_{ab,\frac{N}{2}-1}(u)\dots \mathcal{L}_{ab,2}(u)\mathcal{L}_{ab,1}(u),
    \label{transferrange3}
\end{equation}
where the sites $a,b$ serve as auxiliary spaces. After the gluing, the length of the spin chain is $N/2$.

Expanding the logarithm of the transfer matrix in the spectral parameter, the conserved charges appear as the coefficients. Hence, higher-order logarithmic derivatives of the transfer matrix yield the explicit expressions of these charges. In particular,
\begin{align}
    &\mathcal{Q}_3=t^{-1}(u)t'(u)|_{u=0},\label{Q3fromt}\\
    &\mathcal{Q}_5=t^{-1}(u)t''(u)|_{u=0}-\mathcal{Q}_3^2\label{Q5fromt},
\end{align}
from which one obtains
\begin{align}
    &\mathcal{Q}_3=\sum_{i=1}^{N/2} \mathrm{h}_i,\label{Q3fromh}\\
    &\mathcal{Q}_5=\sum_{i=1}^{N/2} -[\mathrm{h}_{i},\mathrm{h}_{i+1}+\mathrm{h}_{i+2}]-\mathrm{h}_{i}^2+\check{\mathcal{L}}''_{i}\label{Q5fromh},
\end{align}

by considering $\mathcal{L}_{ab,j}(0)=P_{aj}P_{bj}$ and $\mathcal{L}_{ab,j}'(0)=P_{aj}P_{bj}\mathrm{h}_{a,b,j}$. For brevity, we denote $A_{i,i+1,i+2}=A_i$ and impose periodic boundary conditions $A_{{N/2}-1}\equiv A_{{N/2}-1,{N/2},1}$ and $A_{{N/2}}\equiv A_{N/2,1,2}$.

We apply the idea of \cite{de2023lifting,Gombor:2022lco}, which provides an algorithm to detect Yang-Baxter integrability, we write $\mathcal{Q}_5$ as 
\begin{align}
    &\mathcal{Q}_5=\sum_{i=1}^{N/2}\,-[\mathrm{h}_{i},\mathrm{h}_{i+1}+\mathrm{h}_{i+2}]+\tilde{\mathrm{h}}_{i},\label{Q5standardspace}
\end{align}
and claim that $\mathcal{Q}_3$ belongs to a tower of conserved charges if one can find a solution $\tilde{\mathrm{h}}_{i}$ (acting on the local Hilbert space $\mathbb{C}^4$) satisfying
\begin{align}
    &[\mathcal{Q}_5,\mathcal{Q}_3]=0.
\end{align}
 We find a solution\footnote{This solution is unique. The apparent freedom can be removed by the choice of orthonormality among the charges (i.e., ($\mathcal{Q}_i,\mathcal{Q}_j)=2^{-N}\text{tr} (\mathcal{Q}_i^{\dagger} \mathcal{Q}_j)=\delta_{ij}$).} for arbitrary values of the deformation parameters $\alpha, \beta, \gamma, \delta$. The explicit expression of $\tilde{\rom{h}}$ is very long and is provided in \cite{mathematicanotebook}, (\texttt{BulkIntegrability.nb}). Furthermore, in \cite{sajat-medium}, the authors observed that \( \tilde{\mathrm{h}} = 0 \) in all their models and raised the question of whether this vanishing is a general feature or perhaps a consequence of the underlying \( U(1) \) symmetry. In our case, we find \( \tilde{\mathrm{h}} \neq 0 \), suggesting that this property may indeed be directly linked to \( U(1) \) invariance. However, our model also exhibits a $U(1)$-like symmetry, given in \eqref{U1likecharge}, which modifies this correspondence.

\subsubsection*{2.2 Checking conservation of $\mathcal Q_5$.}

Up to this point, we have constructed the charge $\mathcal{Q}_5$ starting from $\mathcal{Q}_3$. We then verified using straightforward computer algebra that this charge commutes with the circuit evolution, i.e.,
\begin{align}
    [\mathbb{U}, \mathcal{Q}_5] = 0. \label{inductionstep}
\end{align}
We have also checked numerically that, by increasing the range of the density of the targeted charge, no additional charges with range between 6 and 10\footnote{except for the spatially reflected charge $Q_7$ described in point 1 of Sec. \ref{evidenceofintegrability1}.} commute with $\mathbb{U}$.

In the next step, we show that there exists an infinite family of charges commuting with both $\mathcal{Q}_3$ and $\mathcal{Q}_5$. To establish a complete proof of integrability for the deformed RCA54 model, it is necessary to demonstrate that these charges also commute with $\mathbb{U}$. While a general proof for arbitrary deformation parameters is currently unavailable, such a proof can be carried out for any fixed numerical choice of $\alpha, \beta, \gamma, \delta$. For variable parameters, we have analytically derived the next conserved charge $\mathcal{Q}_7$ and verified its commutation with $\mathbb{U}$.

\subsubsection*{3. Further gluing and the charge $Q_{14}$}

We consider a nearest-neighbour model obtained from the next-to-nearest neighbour (n.n.n.) by applying the gluing procedure once more. In this construction, the local Hilbert space becomes $\mathbb{C}^{4} \otimes \mathbb{C}^{4} = \mathbb{C}^{16}$. We denote the conserved charges of this model by $\Theta$ and their corresponding densities by $\theta$, while the nearest-neighbor charge density is denoted by $\eta$. 
The densities of the n.n. model are related to those of the n.n.n. model. For instance, under the gluing procedure, we have
\begin{align}
&\eta_{i}=\mathrm{h}_{2i-1}+\mathrm{h}_{2i} \label{etahh},
\end{align}
where, referring to the local spaces in the single-glued picture, this reads explicitly as
$\eta_{12,34}=\mathrm{h}_{123}+\mathrm{h}_{234}$.

For n.n. models, the boost operator is given by \eqref{booststandard}, and in the double-glued space it takes the form
\begin{align}
    &\mathcal{B}[\Theta_2]=\sum_k k\,(\mathrm{h}_{2k-1}+\mathrm{h}_{2k})+\partial_u.
\end{align}
The next conserved charge $\Theta_3$ can be computed as
\begin{align}
    &\Theta_3=[\mathcal{B}[\Theta_2],\Theta_2]=\sum_k \theta^{(3)}_k,\\
&\theta^{(3)}_k=-[\mathrm{h}_{2k-1},\mathrm{h_{2k+1}}]-[\mathrm{h}_{2k-2},\mathrm{h_{2k-1}}+\mathrm{h_{2k}}]+\tilde{\eta}_{k,k+1,k+2,k+3}. \label{theta3glued2}
\end{align}
Since this model originates from the gluing of a range 3 model, we can assume that
\begin{align}
&\theta_{i}^{(3)}=\mathrm{q}_{i}^{(5)}+\mathrm{q}_{i+1}^{(5)},
\end{align}
where $\mathrm{q}_k^{(5)}$ is given in \eqref{Q5standardspace}. By carefully matching these expressions, one observes that in \eqref{theta3glued2} there should be a term $-[\mathrm{h}_{2k-1}, \mathrm{h}_{2k}]$. This term can be absorbed into $\tilde{\eta}_{k,k+1,k+2,k+3}$. Hence, the price of the gluing procedure is that, even for simple models, the term $\tilde{\eta}$ cannot vanish.

By checking the commutativity of the charges before and after the second gluing, we are now in a position to apply the recently proved theorem by  A. Hokkyo \cite{hokkyo2025integrability}:

\vspace{0.2cm}

\textbf{Theorem 1.} 
Let $\Theta_{2}$ and $\Theta_{3}$ be translationally invariant operators satisfying $[\Theta_{2},\Theta_{3}]=0$, and define recursively
\begin{align}
    \Theta_{n+1} = \big[\mathcal{B}[\Theta_{2}], \Theta_n\big], \quad n=2,3,\ldots.
\end{align}
Then, for all $n>2$, it follows that $[\Theta_{2},\Theta_n] = 0$, and more generally
\begin{align}
    [\Theta_m, \Theta_n] = 0, \quad \forall\, m,n > 2.
\end{align}

\vspace{0.2cm}

Since our goal is to establish the integrability of the deformed RCA54 model, and in Sec. \ref{analyticsketch} we provided a proof for a specific numerical choice of the deformation parameters, we now formulate the following conjecture.

\vspace{0.2cm}

\textbf{Conjecture.} Let $\mathbb{U}$ denote the dynamical evolution operator of the deformed RCA54 model. Suppose that
\begin{align}
    [\Theta_2, \mathbb{U}] = [\Theta_3, \mathbb{U}] = 0, 
    \quad \text{with} \quad \Theta_3 = [\mathcal{B}[\Theta_2], \Theta_2].
\end{align}
Then all higher charges $\Theta_n$ generated recursively via
\begin{align}
    \Theta_{n+1} = [\mathcal{B}[\Theta_2], \Theta_n], \quad n>2,
\end{align}
are also conserved, i.e.,
\begin{align}
    [\mathbb{U}, \Theta_n] = 0, \quad \forall\, n>3.
\end{align}

\vspace{0.2cm}

A general proof of this conjecture for variable deformation parameters is currently unavailable. Therefore, it is of interest to explicitly construct the next commuting charge and verify its conservation.

One direct approach to obtain the expression of $\mathcal{Q}_7$ (also denoted $Q_{14}$) is to consider the third logarithmic derivative of the transfer matrix \eqref{transferrange3}:
\begin{align}
    \mathcal{Q}_7 = t^{-1}(u) \, t'''(u)\Big|_{u=0} - 2 \mathcal{Q}_3 \mathcal{Q}_5 - \mathcal{Q}_5 \mathcal{Q}_3 - \mathcal{Q}_3^3. \label{Q7terms}
\end{align}

However, the calculation is long and cumbersome, so we continue to work in the double glued Hilbert spaces. In this picture, the next conserved charge is referred to as $\Theta_4$ and, as before, we can assume that its density can be written as
\begin{align}
&\theta_{i}^{(4)}=\mathrm{q}_{i}^{(7)}+\mathrm{q}_{i+1}^{(7)}\label{definitionoftheta4}.
\end{align}
The charges can now be generated using the  transfer matrix (or equivalently the boost operator). We denote this transfer matrix by $t_\rom{n.n.}(u)$
\begin{align}
    &t_\rom{n.n.}(u)=\tr_A R_{A,{N/4}}(u,\theta_{N/4})R_{A,{N/4}-1}(u,\theta_{{N/4}-1})\dots R_{A,2}(u,\theta_2)R_{A,1}(u,\theta_1),
    \label{transferNNglue}
\end{align}
where $A=\{a,b\}$ and $1,2,\dots,N/4$ label pairs of spaces. This $R$-matrix is related to the range 3 Lax operator\footnote{See conjecture 4 of \cite{sajat-medium}.} as
\begin{align}
    &\check{R}_{A,B}(u,0)=P_{A,B}\, R_{A,B}(u,0)=\check{R}_{ab,cd}(u,0)=\check{\mathcal{L}}_{abc}(u)\check{\mathcal{L}}_{bcd}(u)
    \label{decompositionofR},
\end{align}
where $P_{A,B}$ is the permutation operator exchanging the spaces $A$ and $B$.

By taking the logarithmic derivative of \eqref{transferNNglue}, one obtains
\begin{align}
    &\Theta_2=\sum_{i} \eta_{i}\\
    &\Theta_3=\sum_{i}\theta_{i}^{(3)}=\sum_{i} \,-[\eta_{i},\eta_{{i}+1}]-\eta_{i}^2+\check{{R}}_{i}''(0)=\sum_{i} \,-[\eta_{i},\eta_{{i}+1}]+\tilde{\eta}_i\\
    &\Theta_4=\sum_i -2[\eta_{i},[\eta_{{i}-2},\eta_{{i}-1}]-\tilde{\eta}_{i-1}]-[\eta_{i-1},[\eta_{i-2},\eta_{i-1}]+\tilde{\eta}_{i}]-\tilde{\tilde{\eta}}_{i},
    \label{Q4forNNbig}
\end{align}
where $\tilde{\tilde{\eta}}_{i}$ is an unknown range 2 operator\footnote{We used the expression of $\Theta_4$ from \cite{HbbBoost} and verified its validity in a test model using Mathematica. The expression for $\tilde{\tilde{\eta}}$ is known, but not written in closed form, since it depends on a range-2 operator satisfying the Reshetikhin condition.} acting on $\mathbb{C}^{16}$, i.e., a $256 \times 256$ matrix.

Substituting the expression of $\eta$ from \eqref{etahh} into \eqref{Q4forNNbig}, we attempted to identify the two terms in equation \eqref{definitionoftheta4}. After a lengthy but straightforward calculation, we obtained an explicit expression for $\mathcal{Q}_7$, which still depends on certain coefficients that need to be fixed. These coefficients were determined in a simple test model using \textit{Mathematica} by imposing the commutation relations
\[
[\Theta_4, \Theta_3] = [\Theta_4, \Theta_2] = 0.
\]
We verified the validity of this expression for several test models.

We obtained
\begin{equation}
    \mathrm{q}_{[1,7]}^{(7)}=\Big[\mathrm{h}_5+\mathrm{h}_4+\frac{1}{2}\mathrm{h}_3,[\mathrm{h}_1+\mathrm{h}_2,\mathrm{h}_3]\Big]-[\mathrm{h}_5,\tilde{\mathrm{h}}_3+\tilde{\mathrm{h}}_4]+\frac{1}{2}[\mathrm{h}_3+\mathrm{h}_4,\tilde{\mathrm{h}}_5]+\tilde{\tilde{\mathrm{h}}}_1,\label{Q7standardspace}
\end{equation}
where $\tilde{\mathrm{h}}_i$ is the operator obtained in \eqref{Q5standardspace}, and $\tilde{\tilde{\mathrm{h}}}_i$ is a range-3 operator acting on $\mathbb{C}^4$. 
 Following the same logic as before, we solve
\begin{equation}
[\mathcal{Q}_7,\mathcal{Q}_3]=[\mathcal{Q}_7,\mathcal{Q}_5]=0,
\end{equation}
for the unknown range 3 operator $\tilde{\tilde{\mathrm{h}}}_{i}$. Due to its length, the explicit expression is not reported here, but is provided in \cite{mathematicanotebook}, (\texttt{BulkIntegrability.nb}).

\subsubsection*{3.1 Check that $[\mathbb{U},Q_{14}]=0$.}

We have explicitly verified using symbolic computer algebra that this charge is conserved, i.e.,
\begin{align}
    [\mathbb{U}, Q_{14}] = 0.
\end{align}
As before, we have also checked numerically that, after $Q_{10}$, the next nontrivial conserved charge\footnote{As with $Q_7$, there is an additional charge $Q_{11}$ with range-11 density $q_{11} = 1_2 \otimes \tilde{q}_{10}$, where $\tilde{q}_{10}$ is obtained from $q_{10}$ by reflecting the spatial indices. Since this charge is related to $q_{10}$ by spatial reflection, we do not consider it further in this work.} commuting with $\mathbb{U}$ has range 14; that is, it is $Q_{14}$ up to the addition of lower-range charges.

At this point, we have constructed the first three commuting charges. Using the proof of the star-triangle hypothesis, we can conclude that there exists an infinite family of commuting charges. We have explicitly checked that these three charges are conserved. In the final step (Step 4), we show how this construction corresponds to the perturbative construction of the Lax operator up to order $u^3$. In Sec. \ref{analyticsketch}, we chose specific deformation parameters and constructed the Lax operator, which allowed us to prove the conservation of all the charges.

\subsubsection*{4. From the conserved charges to the perturbative expression of the Lax matrix}

Considering the expressions of the charges \eqref{Q5standardspace} and \eqref{Q7standardspace}, it is clear that the term responsible for the given interaction range is the nested commutator of the lowest-order charge densities $\mathrm{h}_i$. Contributions from higher derivatives of the Lax matrix appear in a separate term, whose interaction range coincides with that of the lowest nontrivial conserved charge—range 3 in this model (with respect to the local space $\mathbb{C}^4$).

If we write the Lax operator $\mathcal{L}_{ab,j}(u)$ as
\begin{equation}
    \check{\mathcal{L}}_{ab,j}(u)=P_{bj}P_{aj}\mathcal{L}_{ab,j}(u)=1+\sum_{i=1}\frac{u^{i}}{i!}\partial_u^{(i)}\check{\La}_{ab,j}(u)|_{u=0},
\end{equation}
we can derive the expressions of the derivatives $\partial_u^{(i)} \check{\mathcal{L}}_{ab,j}(0)$ for $i=1,2,3$ from the charges $\mathcal{Q}_3, \mathcal{Q}_5, \mathcal{Q}_7$, in particular from the range-3 contributions of each charge.

From the boundary condition, we obtain
\begin{align}
    &\check{\mathcal{L}}'_{i}(0)=\mathrm{h}_{i},
\end{align}
by comparing \eqref{Q5fromh} and \eqref{Q5standardspace}
\begin{align}
    &\check{\mathcal{L}}''_{i}(0)=\tilde{\mathrm{h}}_{i}+\mathrm{h}_{i}^2.\label{secondderivative}
\end{align}
From \eqref{Q7terms} and \eqref{Q7standardspace}, the range 3 are
\begin{equation}
    \tilde{\tilde{\rom{h}}}_i=\frac{1}{2}\mathrm{h}_{i}^3+\tilde{\mathrm{h}}_{i}\mathrm{h}_{i}+\frac{1}{2}\mathrm{h}_{i}\tilde{\mathrm{h}}_{i}-\frac{1}{2}\check{\mathcal{L}_{i}}'''(0).\label{thirdderivative}
\end{equation}

Having obtained the expressions for $\tilde{\mathrm{h}}_i$ and $\tilde{\tilde{\mathrm{h}}}_i$ for the deformed RCA54 model, we can extract the perturbative expansion of the Lax operator up to third order in $u$

\begin{equation}
    \check{\mathcal{L}}_{i}(u)=1+u \mathrm{h}_{i}+\frac{u^2}{2}(\tilde{\mathrm{h}}_{i}+\mathrm{h}_{i}^2)+\frac{u^3}{3!}(\mathrm{h}_{i}^3+\mathrm{h}_{i}\tilde{\mathrm{h}}_{i}+2\tilde{\mathrm{h}}_{i}\mathrm{h}_{i}-2\tilde{\tilde{\mathrm{h}}}_{i})+O(u^4), \label{ansatzL}
\end{equation}
with the explicit form of Eq. \eqref{ansatzL} also reported in \cite{mathematicanotebook}, (\texttt{BulkIntegrability.nb}).

We remark that this perturbative expression of the Lax operator, as well as the conserved charges $Q_6$, $Q_{10}$, and $Q_{14}$, are valid for arbitrary values of the deformation parameters.

\subsubsection{Nonperturbative analytical expression of the Lax operator of \ref{analyticsketch} for fixed deformation parameters}
\label{technicaldetailsLax}

Here we provide further clarification on the steps given in \ref{analyticsketch}, to obtain the analytical expression of the Lax operator.

\begin{itemize}
    \item [1.] \textit{Ansatz for the Lax operator.} An important step for the algorithm is choosing an ansatz for the non-zero entries of the Lax operator. As explained in the previous section, each order $n$ in the expansion of the Lax operator in the spectral parameter $u$ contributes to the range 6 part of the corresponding conserved charge $Q_{4n+2}$. For instance, $\tilde{\rom{h}}$ (associated with $\partial_u^2 \mathcal{L}$) appears in $Q_{10}$, while $\tilde{\tilde{\rom{h}}}$ (associated with $\partial_u^3 \mathcal{L}$) appears in $Q_{14}$. These terms are not uniquely fixed, since under periodic boundary conditions one may always add any total divergence (an operator minus its two-site translation). We fix this freedom by choosing the divergence terms such that $\tilde{\rom{h}}$ and $\tilde{\tilde{\rom{h}}}$ are diagonal in the first and last qubit, and so as to minimize the number of non-zero entries in $\rom{h}$.

Furthermore, we notice that at orders $u^2$ and $u^3$, the matrices with the largest number of non-zero entries arise  from $\rom{h}_i^2$ and $\rom{h}_i^3$. We assume that this pattern holds at any order $n$ in $u$: the non-zero entries originate from $\mathrm{h}_i^n$. By examining the non-zero elements of $\mathrm{h}_i^n$ for integer $n$, we observe that considering terms up to $\mathrm{h}_i^4$ is sufficient to capture all non-zero entries in $\mathrm{h}_i^n$ for $n>4$. This leaves us with an ansatz containing 248 non-zero entries, which we label as $f_k(u)$ with $k=1,\dots,248$. This ansatz is displayed in Fig.~\ref{nonzeroLax}.

\item[2.] \textit{Solve $[t(u),Q_6]=0$ perturbatively.} Since we want the transfer matrix to generate the commuting charges, it must commute with $Q_6$ at all orders in $u$. We solve this condition up to the 35th order in perturbation theory, fixing the chain length at $N=8$ for each order.

We are aware that using a small $N$ could introduce periodic wrapping effects for long operators, but in this case it does not pose a problem. As seen in \eqref{Q5fromh} and \eqref{Q7standardspace}, the range of a charge is determined by the nested commutators, while the higher-order derivatives of the Lax operator always appear in terms with the minimal range of the conserved charge (6 in this case). Therefore, the wrapping effects resolve themselves automatically, and it is sufficient to take $N = 6 + 2$. Note that $N$ must be even, since we glue the chain once and treat $Q_6$ as the translationally invariant charge $\mathcal{Q}_3$.
\item[3.] \textit{Analytical expression of the entries of the Lax operator}

We aim to find explicit analytical expressions --- functions of $u$ --- for all entries of the Lax operator.
\begin{itemize}
    \item [a.] We observe that 43 of the entries are polynomials of degree 4 or lower. Since we performed the perturbative expansion up to the 35th order, we assume that these entries do not receive higher-order corrections. This suggests that the remaining entries might also be expressible as ratios of polynomials, with numerator of degree at most 4.

Using the \texttt{PadeApproximant} of Mathematica, we were able to resum 11 of these entries. Their expressions take the form
\begin{align}
    f_i(u) = \frac{P_4(u)}{P_2(u)},
\end{align}
where $P_k(u)$ denotes a polynomial of degree at most $k$.
    \item [b.] In the same spirit, we also consider ratios between different entries. In this way, 126 entries can be fixed as functions of the remaining ones.
    \item[c.]
Setting $N=2,4$, we constructed two transfer matrices: one obtained from the perturbative expansion of the Lax operator, and the other from the ansatz given in Fig \ref{nonzeroLax}. This ansatz can be further simplified because, from the perturbative expressions, some entries are identical; after accounting for this, 229 entries remain to be fixed. We also include the known entries determined by points a and b. Requiring the entries of the two transfer matrices to coincide allows us to fix 13 additional entries.
\item[d.] We fix the remaining entries by imposing, for $N=8$, the commutation relation $[t(u),Q_6]=0$. For one of the entries, we find that it involves a square root:
\begin{align}
    f_{120}(u)= \frac{W_4(u)\left(W_2(u)+ \sqrt{G_4(u)}\right)}{F_4(u)} \, ,\label{f120entries}
\end{align}
where $W_i(u)$, $G_i(u)$, and $F_i(u)$ are polynomials of degree $i$. Although one can remove the square root by introducing an elliptic parametrization of $u$, the resulting solution involves Weierstrass elliptic functions and is not simpler than the original expression. For this reason, we keep the original parametrization.

In this way, we have fixed all remaining entries, obtaining the analytic dependence of the Lax matrix elements, which are all rational functions of $u$ possibly containing a single square root. The explicit expressions of all entries as functions of $f_{120}(u)$, as well as the explicit form of $f_{120}(u)$ itself, can be found in \cite{mathematicanotebook}, (\texttt{BulkIntegrability.nb}).

\end{itemize}
This provides the exact expression of the Lax operator for a fixed choice \eqref{numericalchoice} of the deformation parameters.

Furthermore, we have explicitly verified that, for $N=8$ and $N=10$, the commutation relation $[t(u),\mathbb{U}]=0$ holds. In the following, we provide an algebraic proof of this statement, thereby establishing its validity for arbitrary system size $N$. A graphical version of the proof was presented in Figs.~\ref{Rllproof} and \ref{tuproof}.
\end{itemize}
\subsubsection{Algebraic proof of $[t(u),t(v)]=0$}

The transfer matrix $t(u)$ is defined in \eqref{transferrange3}. Our goal is to prove that $[t(u),t(v)]=0.$

We consider
\begin{align}
    t(v)t(u)=& \, \text{tr}_{a,b,c,d} \tilde{\mathcal{L}}_{ab,\frac{N}{2}}\tilde{\mathcal{L}}_{ab,\frac{N}{2}-1}\dots \tilde{\mathcal{L}}_{ab,2}\tilde{\mathcal{L}}_{ab,1}\mathcal{L}_{cd,\frac{N}{2}}\mathcal{L}_{cd,\frac{N}{2}-1}\ \mathcal{L}_{cd,2}\mathcal{L}_{cd,1}=\nonumber\\
    & \, \text{tr}_{a,b,c,d} \tilde{\mathcal{L}}_{ab,\frac{N}{2}}\mathcal{L}_{cd,\frac{N}{2}}\tilde{\mathcal{L}}_{ab,\frac{N}{2}-1}\mathcal{L}_{cd,\frac{N}{2}-1}\dots \tilde{\mathcal{L}}_{ab,2}\mathcal{L}_{cd,2} \tilde{\mathcal{L}}_{ab,1}\mathcal{L}_{cd,1}=\nonumber\\
    &\text{tr}_{a,b,c,d} {R}_{ab,cd}\tilde{\mathcal{L}}_{ab,\frac{N}{2}}\mathcal{L}_{cd,\frac{N}{2}}\tilde{\mathcal{L}}_{ab,\frac{N}{2}-1}\mathcal{L}_{cd,\frac{N}{2}-1}\dots \tilde{\mathcal{L}}_{ab,2}\mathcal{L}_{cd,2} \tilde{\mathcal{L}}_{ab,1}\mathcal{L}_{cd,1}{R}_{ab,cd}^{-1}=\nonumber\\
    &\text{tr}_{a,b,c,d} \mathcal{L}_{cd,\frac{N}{2}}\tilde{\mathcal{L}}_{ab,\frac{N}{2}}{R}_{ab,cd}\tilde{\mathcal{L}}_{ab,\frac{N}{2}-1}\mathcal{L}_{cd,\frac{N}{2}-1}\dots \tilde{\mathcal{L}}_{ab,2}\mathcal{L}_{cd,2} \tilde{\mathcal{L}}_{ab,1}\mathcal{L}_{cd,1}{R}_{ab,cd}^{-1}=\dots=\nonumber\\
&\text{tr}_{a,b,c,d}\mathcal{L}_{cd,\frac{N}{2}}\tilde{\mathcal{L}}_{ab,\frac{N}{2}}\mathcal{L}_{cd,\frac{N}{2}-1}\tilde{\mathcal{L}}_{ab,\frac{N}{2}-1}\dots \mathcal{L}_{cd,2}\tilde{\mathcal{L}}_{ab,2} \mathcal{L}_{cd,1}\tilde{\mathcal{L}}_{ab,1}=\nonumber\\
&\text{tr}_{a,b,c,d}\mathcal{L}_{cd,\frac{N}{2}}\mathcal{L}_{cd,\frac{N}{2}-1}\dots \mathcal{L}_{cd,2}\mathcal{L}_{cd,1}\tilde{\mathcal{L}}_{ab,\frac{N}{2}}\tilde{\mathcal{L}}_{ab,\frac{N}{2}-1}\dots \tilde{\mathcal{L}}_{ab,2}\tilde{\mathcal{L}}_{ab,1}= t(u)t(v),
\end{align}
where, for brevity, we introduce the notation $\mathcal{L}=\mathcal{L}(u)$, $\tilde{\mathcal{L}}=\mathcal{L}(v)$ and $R_{ab,cd}=R_{ab,cd}(v,u)$ and we assume the existence of an invertible operator $R_{ab,cd}$ that solves the RLL relation \eqref{RLLmediummodel}. Explicitly,
\begin{align}
    &R_{ab,cd}(v,u)\mathcal{L}_{ab,j}(v)\mathcal{L}_{cd,j}(u)=\mathcal{L}_{cd,j}(u)\mathcal{L}_{ab,j}(v)R_{ab,cd}(v,u).
    \label{RLLmediummodelspec}
\end{align}
Since $R$ is a matrix of dimension $2^8 \times 2^8$, we could check its existence and invertibility only for fixed choices of $u$ and $v$, specifically setting $u = 1/6$ and $v = 2/5$.

By considering $\mathcal{L}_{a,b,j}=P_{a,j}P_{b,j}\check{\mathcal{L}}_{a,b,j}$, ${R}_{ab,cd}=P_{a,c}P_{b,d}\check{{R}}_{ab,cd}$, we can rewrite \eqref{RLLmediummodelspec} in the original unglued space $\mathbb{C}^2$ as \eqref{RLLcheck}. We remark that the operators $\check{R}$ and  $\check{\mathcal{L}}$ are diagonal in the first and last space they act upon. A graphical representation of \eqref{RLLcheck} and the corresponding proof of $[t(u),t(v)] = 0$ are provided in Sec.~\ref{prooftutv}.

\label{algebraicprooftutv}

\subsubsection{Algebraic proof  of $[t(u),\mathbb{U}]=0$.}

\label{algebraicprooftuU}
In this section, we provide an algebraic proof that $[t(u),\mathbb{U}]=0$, where $t(u)$ is defined in \eqref{transferrange3} and $\mathbb{U}$ in \eqref{Ucircuit}. For convenience, we recall their explicit expressions

\begin{equation}
    t(u)=\text{tr}_{a,b} \mathcal{L}_{ab,\frac{N}{2}}(u)\mathcal{L}_{ab,\frac{N}{2}-1}(u)\dots \mathcal{L}_{ab,2}(u)\mathcal{L}_{ab,1}(u),
\end{equation}
and
\begin{equation}
    \mathbb{U}=\big(U_{123}U_{345}U_{567}\dots U_{N-1,N,1}\big)\big(U_{234}U_{456}U_{678}\dots U_{N,1,2}\big).
\end{equation}
Using the fact that operators acting on different spaces commute, we can rewrite $\mathbb{U}$ as
\begin{equation}
    \mathbb{U}=\big(U_{N-1,N,1}U_{N,1,2}\big)\dots \big(U_{345}U_{456} \big)\big(U_{123}U_{234} \big).
\end{equation}
We now consider the glued picture as in Point 2 of Sec.~\ref{perturbativesketch} and define an operator $D$ acting on two glued spaces
\begin{equation}
    \check{D}_{A,B}=U_{a_1,a_2,b_1}U_{a_2,b_1,b_2},\,\,\,\,\, D_{A,B}=P_{A,B} \check{D}_{A,B},
\end{equation}
where $A=\{a_1,a_2\}$, $B=\{b_1,b_2\}$.
In this way, we can write
\begin{equation}
\mathbb{U}= \mathbb{P} \,\tilde{\mathbb{U}}, 
\end{equation}
where $\mathbb{P}$ is the cyclic permutation,
\begin{align}
    &\mathbb{P}=P_{\frac{N}{2}-1,\frac{N}{2}}P_{\frac{N}{2}-2,\frac{N}{2}}\dots P_{2,\frac{N}{2}}P_{1,\frac{N}{2}},\label{permC4}\\
    &\tilde{\mathbb{U}}=\text{tr}_a D_{a,\frac{N}{2}}D_{a,\frac{N}{2}-1}\dots D_{a,1}.
\end{align}

We remark that, in the glued picture, space indices refer to $\mathbb{C}^4$, so that $\mathbb{P}$ corresponds to a cyclic permutation implementing a shift by two sites in the original $\mathbb{C}^2$ space. Explicitly, $P_{1,\frac{N}{2}}$ of \eqref{permC4} is the permutation operator swapping the glued sites $1$ and $\frac{N}{2}$ and can be written in term of the original space $\mathbb{C}^2$ as $P_{1,N-1}P_{2,N}$.

Since $t(u)$ is translationally invariant, it is sufficient to prove that
\begin{equation}
    [t(u),\tilde{\mathbb{U}}]=0.
\end{equation}

We now search for an intertwiner relation. For brevity, we omit the $u$-dependence in $\mathcal{L}$
\begin{align}
t(u)\,\tilde{{\mathbb{U}}}\,=\,&\text{tr}_{a,b,c}\mathcal{L}_{a,b,\frac{N}{2}}\mathcal{L}_{a,b,\frac{N}{2}-1}\mathcal{L}_{a,b,\frac{N}{2}-2}\dots \mathcal{L}_{a,b,1}D_{c,\frac{N}{2}}D_{c,\frac{N}{2}-1}\dots D_{c,1}=\\
&\text{tr}_{a,b,c}A_{a,b,c}\mathcal{L}_{a,b,\frac{N}{2}}D_{c,\frac{N}{2}}\mathcal{L}_{a,b,\frac{N}{2}-1}D_{c,\frac{N}{2}-1}\dots \mathcal{L}_{a,b,1} D_{c,1}A_{a,b,c}^{-1},
\end{align}
where we assume the existence of an invertible operator $A_{a,b,c}$  satisfying
\begin{align}
&A_{a,b,c}\mathcal{L}_{a,b,j}D_{c,j}=D_{c,j}\mathcal{L}_{a,b,j}A_{a,b,c}. \label{ALuu}
\end{align}
In this way, we can repeatedly apply the intertwiner relation \eqref{ALuu} to obtain

\begin{align}
    &\text{tr}_{a,b,c}D_{c,\frac{N}{2}}\mathcal{L}_{a,b,\frac{N}{2}}A_{a,b,c}\mathcal{L}_{a,b,\frac{N}{2}-1}D_{c,\frac{N}{2}-1}\dots \mathcal{L}_{a,b,1}D_{c,1}A_{a,b,c}^{-1}=\dots=\\
    &\text{tr}_{a,b,c}D_{c,\frac{N}{2}}\mathcal{L}_{a,b,\frac{N}{2}}D_{c,\frac{N}{2}-1}\mathcal{L}_{a,b,\frac{N}{2}-1}\dots D_{c,1}\mathcal{L}_{a,b,1}=\\
    &\text{tr}_{a,b,c}D_{c,\frac{N}{2}}D_{c,\frac{N}{2}-1}\dots D_{c,1} \mathcal{L}_{a,b,\frac{N}{2}}\mathcal{L}_{a,b,\frac{N}{2}-1}\dots \mathcal{L}_{a,b,1}=\tilde{\mathbb{U}}\,t(u).
\end{align}

We set $u = 1/6$ and solve \eqref{ALuu} for $A_{a,b,c}$, verifying that it is indeed invertible.

By considering, $\mathcal{L}_{a,b,j}=P_{a,j}P_{b,j}\check{\mathcal{L}}_{a,b,j}$, ${A}_{a,b,j}=P_{a,j}P_{b,j}\check{{A}}_{a,b,j}$, we can rewrite \eqref{ALuu} in the original unglued space $\mathbb{C}^2$ as
\eqref{RLLUUa}. 
We remark that the operator $\check{A}_{[1-6]}$ is diagonal in the first and the last space. A graphical representation of \eqref{RLLUUa} and a corresponding graphical proof of $[t(u),\mathbb{U}]=0$ were provided in Sec. \ref{prooftuU}.

\section{Boundary Driven Case}
\label{boundarydrivness}

We now turn to the integrability of the boundary-driven stochastic setup. The dynamics of the model are described in Sec. \ref{openboundaryconditiondynamics}: in the bulk, we consider the stochastic deformation \eqref{stochasticcond} of the RCA54 model, while at the boundaries, we implement the family of baths known as conditional driving, defined in \eqref{conditionaldriving}. In this setting, \textit{integrability} refers to the ability to exactly solve for (or explicitly construct) the non-equilibrium steady state (NESS), that is, to obtain a closed-form analytical expression using a suitable tensor-network representation (e.g., matrix product states or patch-state~\cite{prosen2016integrability}).

\subsection{The Non-Equilibrium Steady State}
\label{nesssection}
The NESS probability vector $\mathbf{p}$ is a fixed point of the markovian dynamical map
\begin{align}
&\mathbb{U}\,\mathbf{p}=\mathbf{p}.
\end{align}
First, in Sec. \ref{uniqueness}, we discuss the existence and uniqueness of NESS, and  then in Sec. \ref{patch} we provide its explicit construction.
\subsubsection{Strong ergodicity and uniqueness of NESS}
\label{uniqueness}
We now establish the irreducibility and aperiodicity of Markov chain generated by the many-body stochastic matrix $\mathbb{U}$ (\ref{bigUopen}), which, according to the Perron--Frobenius theorem, implies the existence of a unique NESS and the asymptotic convergence of any initial state to it (ergodicity and dynamical mixing, sometimes referred to as strong ergodicity \cite{levin2008markov}).

\paragraph*{Theorem.} The $2^N\times 2^N$ matrix $\mathbb{U}$ given in \eqref{bigUopen} is irreducible and aperiodic for an open set of the boundary parameters $0<a,b,c,d<1$ of \eqref{conditionaldriving} and a set of the bulk parameters $0\le\beta,\gamma < 1$.
\paragraph*{Proof.}

In \cite{prosen2016integrability}, the theorem was proven for the case where the bulk consists of the undeformed RCA54 model ($\gamma=\beta=0$). Here, we follow essentially the same idea to generalize that result.

The proof of \textit{irreducibility} amounts to showing that, for any two initial configurations $s, s' \in \{0,1\}^N$, there exists a natural number $t_0$ such that
\begin{align}
    (\mathbb{U}^{t_0})_{s,s'}>0. \label{t0}
\end{align}
Compared to the deterministic case, we now have additional branching in the Markov graph due to the stochastic bulk (except for the case $\beta = \gamma = 1$, where the bulk dynamics reduces to a trivial deterministic all-NOT map, which is excluded in the theorem). Hence, if such a $t_0$ exists in the undeformed case for which \eqref{t0} holds, the same must also hold in the stochastically deformed case.

This can be understood intuitively: in the undeformed case, the bulk evolution is deterministic and only the boundaries introduce stochasticity, so the time required to reach configuration $s'$ from $s$ is $t_0'$. In the present case, the bulk evolution is also stochastic, so more configurations can be reached probabilistically at each step. Therefore, configuration $s'$ can be reached with nonzero probability at the same or an earlier time, i.e.\ $t_0 \le t_0'$.

An irreducible matrix $\mathbb{U}$ is \textit{aperiodic} if, for any configuration $s$, the greatest common divisor of the set of recurrence times $\{t_j\}$ for which \eqref{t0} holds is equal to $1$. Since the possibility of a recurrence cannot disappear when the dynamics becomes more stochastic, the set of recurrence times in the stochastic case contains that of the deterministic case, and therefore their greatest common divisor remains equal to $1$.

\vspace{0.5cm}

In summary, as follows from Perron-Frobenius theorem, the NESS exists, is unique, and is reached asymptotically in time from any initial state. In what follows, we construct  
the probability vector $\mathbf{p}$ in terms of a
hybrid patch -- matrix product ansatz.

\subsubsection{Expression of the NESS: Patch Matrix Product Ansatz}
\label{patch}
Since the propagator $\mathbb{U}$ of \eqref{bigUopen} can be factored into two half-time steps, the NESS is defined as
\begin{align}
&\mathbb{U}\,\mathbf{p}=\mathbb{U}_\mathrm{o}\mathbb{U}_\mathrm{e}\,\mathbf{p}=\mathbf{p},\label{ness1}
\end{align}
where $\mathbb{U}_{\mathrm{e}}$ and $\mathbb{U}_{\mathrm{o}}$ are given in \eqref{bigUeopen} and \eqref{bigUoopen}. The $2^N$ dimensional probability vector $\mathbf{p}$ can be split into even- and odd-time parts, satisfying
\begin{align}    &\mathbb{U}_\mathrm{e}\,\mathbf{p}=\mathbf{p}',\label{ness2}\\
&\mathbb{U}_\mathrm{o}\,\mathbf{p}'=\mathbf{p}.\label{ness3}
\end{align}
In what follows, we use bold characters to indicate vectors over the physical space (of cell configurations), while their components may still be nontrivial operators over some auxiliary vector space ${\mathcal V}_{\rm a}$.
A roman subscripts denote local positions on which this vector-operators act nontrivially.
For example,
$\mathbf{A}_{2,3}=\mathbb{1}\otimes  A_{s_2,s_3}\otimes \mathbb{1}\otimes \dots \otimes \mathbb{1}$, where
$A_\rom{s,s'}\in {\rm End}({\mathcal V}_{\rm a})$, $s,s'\in\{0,1\}$.

\begin{figure}[H]
    \centering    \includegraphics[width=0.4\textwidth]{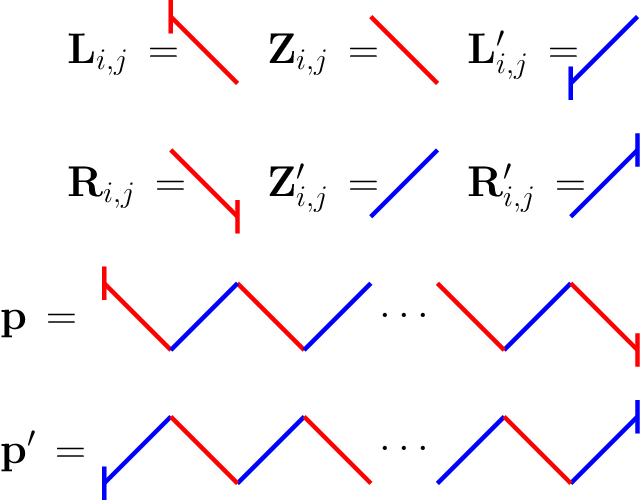}
    \caption{Graphical representation of the matrices $\LL, \ZZ, \LLp, \RR, \ZZp, \RRp$ and the two time slices of NESS, $\mathbf{p}, \mathbf{p}'$.}
    \label{figurematrices}
\end{figure}

We take a staggered patch matrix product ansatz for the NESS probability vectors
\begin{align}
&\mathbf{p}=\LL_{12}\ZZp_{23}\ZZ_{34}\ZZp_{45}\ZZ_{56}\dots \ZZ_{N-3,N-2} \ZZp_{N-2,N-1} \RR_{N-1,N},\\
&\mathbf{p}'=\LLp_{12}\ZZ_{23}\ZZp_{34}\ZZ_{45}\ZZp_{56}\dots \ZZp_{N-3,N-2} \ZZ_{N-2,N-1} \RRp_{N-1,N},
\end{align}
where $\ZZ, \, \ZZp,$ represent two 4-tuples of auxiliary space matrices $Z_{s,s'},Z'_{s,s'}\in{\rm End}({\mathcal V}_{\rm a}) $, while $\LL,\,\LLp$, and $\RR,\,\RRp$ are pairs of $4$-tuples of row and column vectors, respectively, i.e. elements of $\mathcal V_{\rm a}^*$ and $\mathcal V_{\rm a}$. We represent this ansatz graphically in Fig. \ref{figurematrices}.

For this ansatz to represent a fixed point of the many-body Markov chain, the tensors must satisfy a set of boundary and bulk equations, namely patch-state analogs of the Ghoshal–Zamolodchikov relations at the boundaries and of the Zamolodchikov–Faddeev relations in the bulk (in analogy to the vertex case, see e.g.~\cite{vanicat2018integrable}).
In particular, we require that the matrices $\ZZ, \, \ZZp,$ and the vectors $\LL,\,\LLp,\,\RR,\,\RRp$ satisfy the following equations. For the boundaries:
\begin{align}
&U_{123}\LL_{12}\ZZp_{23}=\LLp_{12}\ZZ_{23}\label{ULZpboundary}\\
&U_{12}^R\,\RR_{12}\,=\,\RRp_{12}\label{URboundary}
\end{align}
\begin{align}
&U_{12}^L\LLp_{12}=\LL_{12}\label{ULppboundary}\\    &U_{123}\ZZ_{12}\RRp_{23}=\ZZp_{12}\RR_{23}\label{UZRpboundary}
\end{align}
and for the bulk:
\begin{align}
&U_{123}\ZZ_{12}\ZZp_{23}=\ZZp_{12}\ZZ_{23}.\label{UZZbulk}
\end{align}

\begin{figure}[H]
    \centering    \includegraphics[width=0.4\textwidth]{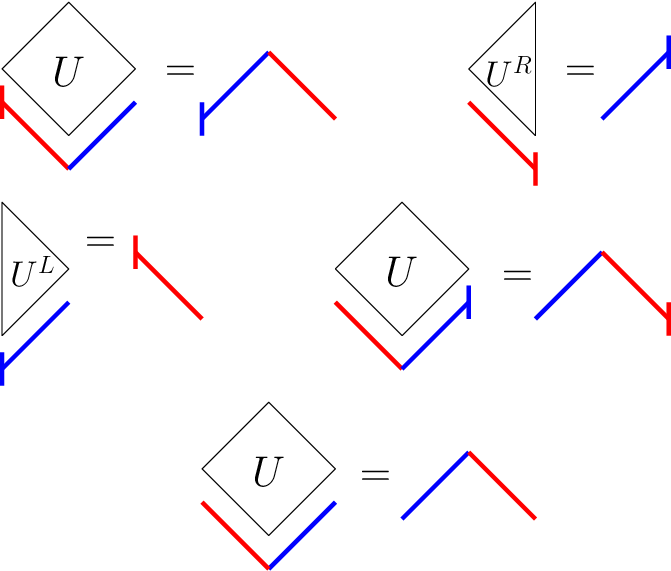}
    \caption{Graphical representation of the equations \eqref{ULZpboundary}-\eqref{UZZbulk}.}
    \label{figureequations}
\end{figure}

\begin{figure}[h]
    \centering    \includegraphics[width=0.5\textwidth]{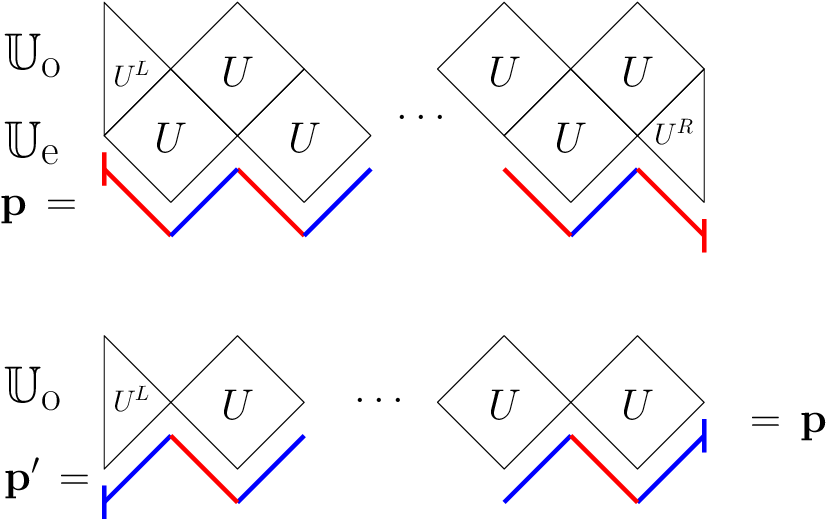}
    \caption{Graphical representation of the stationary point equation \eqref{ness1}.}
    \label{figureNESS}
\end{figure}

\subsubsection{IRF version of Zamolodchikov-Faddeev algebra}
By writing the bulk equation component-wise \eqref{UZZbulk}, one can finds a simple quadratic algebra  which can
be interpreted as a IRF or face analog of Zamolodchikov–Faddeev algebra~\cite{zamolodchikov1979,vanicat2018integrable}
\begin{eqnarray}
    &&(1-\gamma)\, Z_{00}Z'_{00}+\beta \,Z_{01}Z'_{10}=Z'_{00}Z_{00},\nonumber\\
    &&\gamma\, Z_{00}Z'_{00}+(1-\beta) \,Z_{01}Z'_{10}=Z'_{01}Z_{10},\label{qalg}\\
    &&Z_{s,s'}Z'_{s',s''}=Z_{s,\bar{s}'}Z'_{\bar{s}',s''},\,\,\,\,\,\textrm{for}\,\,\,(s,s'')\neq (0,0),
    \nonumber 
\end{eqnarray}
where $\bar{s}\equiv 1-s$.
These are eight quadratic relations among eight operators $Z_{s,s'},Z'_{s,s'}$. The remainder of the section is dedicated to finding its appropriate representation in $\mathcal V_{\rm a}$.

\subsubsection{Representation}
\label{representationboundary}
To obtain an explicit representation of the quadratic 
algebra~(\ref{UZZbulk}) (or (\ref{qalg})) that simultaneously obeys the boundary relations~(\ref{ULZpboundary}--\ref{UZRpboundary}), 
we employ the following structured matrix ansatz for tensors $\ZZ,\ZZp, \LL,\LL',\RR,\RRp$, inspired by heavy and numerous numerical simulations as well as many trial-and-error attempts (until it finally worked!).

Specifically, we found that auxiliary
space should be infinite dimensional and decomposed as a direct sum of 3-dimensional spaces 
\begin{equation}
    \mathcal V_a = \bigoplus_{n=1}^\infty \mathcal V^{(n)},\quad 
    \mathcal V^{(n)}=\mathbb C^3,
    \label{eq:directsum}
\end{equation}
where the integer $n$ is referred to as the {\em level}. The boundary states are assumed to require only level-1 components,  i.e. they represent lowest-weight states of the representation:
\begin{equation}
L_{s,s} = 
\begin{pmatrix}
l^{(0)}_{s,s'}|0\cdots
\end{pmatrix},\;\;
L'_{s,s} = 
\begin{pmatrix}
l^{'(0)}_{s,s'}|0\cdots
\end{pmatrix},\quad
R_{s,s} = 
\begin{pmatrix}
r^{(0)}_{s,s'}|0\cdots
\end{pmatrix}^T,\;\;
R'_{s,s} = 
\begin{pmatrix}
r^{'(0)}_{s,s'}|0\cdots
\end{pmatrix}^T.
\label{BT}
\end{equation}
However, to distinguish the boundary vector components from level-1 
components of the bulk tensors for the purpose of systematically solving boundary\&bulk equations, we refer to the boundary vector components (\ref{BT}) as level-0 variables (labeled by superscript $^{(0)}$).

The bulk operators, however, involve ladder operators for the representation, and can be represented in terms of a block tridiagonal ansatz:
\begin{align}
    &Z_{s,s'}=\sum_{n=1}^{\infty} Z_{s,s'}^{(n,0)}\ket{n}\bra{n}+Z_{s,s'}^{(n+1,+)}\ket{n}\bra{n+1}+Z_{s,s'}^{(n+1,-)}\ket{n+1}\bra{n},\label{Ztridiagblocks}\\        &Z'_{s,s'}=\sum_{n=1}^{\infty} Z_{s,s'}'^{(n,0)}\ket{n}\bra{n}+Z_{s,s'}'^{(n+1,+)}\ket{n}\bra{n+1}+Z_{s,s'}'^{(n+1,-)}\ket{n+1}\bra{n},\label{Zptridiagblocks}
\end{align}
where $Z_{s,s'}^{(n,i)}, Z_{s,s'}'^{(n,i)}$ are $3 \times 3$ matrices labeled with raising/lowering index $i \in {-,0,+}$ and level $n \in \mathbb{N}$. See Fig. \ref{figureZansatz} for an
illustration of their structure.
\begin{figure}[h]
\centering
\[
Z_{s,s'} =
\left(
\begin{tikzpicture}[scale=0.92, baseline=(current bounding box.center)]
    \tikzstyle{block}=[
      draw,
      minimum width=1cm,
      minimum height=1cm,
      inner sep=0pt,
      align=center
    ]
    \node[block] (a11) at (0,0) {$Z^{(1,0)}_{s,s'}$};
    \node[block,right=-0.3pt of a11] (a12) {$Z^{(2,+)}_{s,s'}$};
    \node[block,below=-0.3pt of a11] (a21) {$Z^{(2,-)}_{s,s'}$};
    \node[block,right=-0.3pt of a21] (a22) {$Z^{(2,0)}_{s,s'}$};
    \node[block,right=-0.3pt of a22] (a23) {$Z^{(3,+)}_{s,s'}$};
    \node[block,below=-0.3pt of a22] (a32) {$Z^{(3,-)}_{s,s'}$};
    \node[block,right=-0.3pt of a32] (a33) {$Z^{(3,0)}_{s,s'}$};
    \node at ($(a33)+(1.3,-1)$) {$\cdots$};
   \node[block] (an1) at ($(a11)+(4.5cm,-3.5cm)$) {$Z^{(n,+)}_{s,s'}$};
    \node[block,below=-0.3pt of an1] (an2) {$Z^{(n,0)}_{s,s'}$};
    \node[block,left=-0.3pt of an2] (an3) {$Z^{(n,-)}_{s,s'}$};
    \node at ($(an3)+(1.8,-1)$) {$\cdots$};
\end{tikzpicture}
\right)\]
\vspace{-0.7cm}

\caption{Block tridiagonal structure of each components $s,s'$ of the infinite dimensional matrix $\ZZ$. The structure of $\ZZp$ is the same.}
\label{figureZansatz}
\end{figure}
A more detailed structure of matrices $\ZZ,\ZZp$ and vectors $\LL,\LLp,\RR,\RRp$, in particular their non-zero or trivial entries ($\pm 1$), is presented in Appendix \ref{masksappendix}.
We note that in order to determine any physical property of the NESS (i.e. to calculate any observable) for a finite chain of size $N$, it suffices to consider the representation of the patch matrix product tensors up to level $n = N/2$.

By explicitly implementing the block tri-diagonal ansatz \eqref{Ztridiagblocks}-\eqref{Zptridiagblocks}, 
the bulk equations \eqref{UZZbulk}
reduce to a 3-point recurrence scheme
coupling components across three consecutive blocks, e.g. $n,\,n-1,\,n-2$.
Depending on the `initial' and `final' level (which can be either $n-2,\, n-1,\, n$), and denoting the maximal relevant level as $n$,
all nontrivial equations in the system ~\eqref{UZZbulk} reduce to 5 
distinct $3\times 3$ matrix equations (cf. the definition of the face interaction tensor (\ref{deformation})):  
\begin{align}
&\sum_{t_2=0}^1
(f_{s_1,s_3})_{s_2}^{t_2}
 Z_{s_1,t_2}^{(n,-)}{Z'}_{t_2,s_3}^{(n-1,-)}={Z'}_{s_1,s_2}^{(n,-)}{Z}_{s_2,s_3}^{(n-1,-)},\label{minusminus}\\    
 &\sum_{t_2=0}^1
 (f_{s_1,s_3})_{s_2}^{t_2}
  Z_{s_1,t_2}^{(n-1,+)}{Z'}_{t_2,s_3}^{(n,+)}={Z'}_{s_1,s_2}^{(n-1,+)}{Z}_{s_2,s_3}^{(n,+)}
\label{plusplus},
\end{align}
\begin{align}
&\sum_{t_2=0}^1(f_{s_1,s_3})_{s_2}^{t_2} \big[Z_{s_1,t_2}^{(n-1,0)}{Z'}_{t_2,s_3}^{(n,+)}+Z_{s_1,t_2}^{(n,+)}{Z'}_{t_2,s_3}^{(n,0)}\big]={Z'}_{s_1,s_2}^{(n-1,0)}Z_{s_2,s_3}^{(n,+)}+{Z'}_{s_1,s_2}^{(n,+)}Z_{s_2,s_3}^{(n,0)}\label{zeroplus},\\    &\sum_{t_2=0}^1 (f_{s_1,s_3})_{s_2}^{t_2}
\big[Z_{s_1,t_2}^{(n,-)}{Z'}_{t_2,s_3}^{(n-1,0)}+Z_{s_1,t_2}^{(n,0)}{Z'}_{t_2,s_3}^{(n,-)}\big]={Z'}_{s_1,s_2}^{(n,-)}Z_{s_2,s_3}^{(n-1,0)}+{Z'}_{s_1,s_2}^{(n,0)}Z_{s_2,s_3}^{(n,-)}\label{zerominus},
\end{align}
\begin{align}
\sum_{t_2=0}^1(f_{s_1,s_3})_{s_2}^{t_2}
 &\big[Z_{s_1,t_2}^{(n,+)}{Z'}_{t_2,s_3}^{(n,-)}+Z_{s_1,t_2}^{(n-1,0)}{Z'}_{t_2,s_3}^{(n-1,0)}+Z_{s_1,t_2}^{(n-1,-)}{Z'}_{t_2,s_3}^{(n-1,+)}\big]=\label{plusminus}\\
    &{Z'}_{s_1,s_2}^{(n,+)}{Z}_{s_2,s_3}^{(n,-)}+{Z'}_{s_1,s_2}^{(n-1,0)}{Z}_{s_2,s_3}^{(n-1,0)}+{Z'}_{s_1,s_2}^{(n-1,-)}{Z}_{s_2,s_3}^{(n-1,+)}\nonumber.
\end{align}
Each of these identities can be represented diagrammatically, as illustrated in Fig. \ref{figurelevels}.
In particular, we depict the blocks $Z^{(n,+)}, Z^{(n,-)}$, and $Z^{(n,0)}$ (and similarly for $Z'$) diagrammatically as

\[
\ZZ^{(n,-)}:\,\,\,\, \raisebox{-0.4\height}{\includegraphics[width=0.13\linewidth]{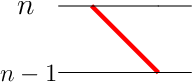}};\,\,\,\,\,\,\ZZ^{(n,0)}:\,\,\,\, \raisebox{-0.4\height}{\includegraphics[width=0.13\linewidth]{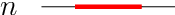}};\,\,\,\,\,\,\ZZ^{(n,+)}:\,\,\,\, \raisebox{-0.4\height}{\includegraphics[width=0.13\linewidth]{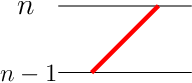}}
\]
\[
\ZZp^{(n,-)}:\,\,\,\, \raisebox{-0.4\height}{\includegraphics[width=0.13\linewidth]{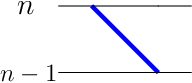}};\,\,\,\,\,\,\ZZp^{(n,0)}:\,\,\,\, \raisebox{-0.4\height}{\includegraphics[width=0.13\linewidth]{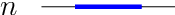}};\,\,\,\,\,\,\ZZp^{(n,+)}:\,\,\,\, \raisebox{-0.4\height}{\includegraphics[width=0.13\linewidth]{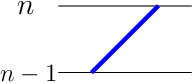}}
\]
Hence, the equations read (note that the face interaction tensor is implemented on the left-hand sides, but not explicitly indicated to lighten notation):
\begin{figure}[h]
    \centering 
    
    \begin{minipage}{0.68\textwidth}
    \includegraphics[width=\linewidth]{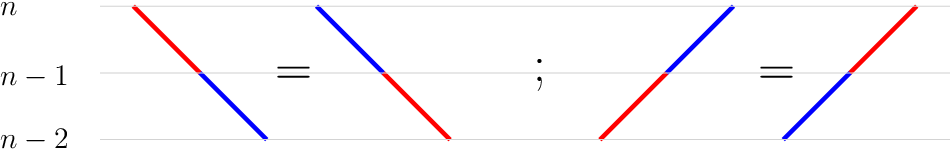}
\end{minipage}
\hfill
\begin{minipage}{0.28\textwidth}
    \small
    Eqs. \eqref{minusminus} and \eqref{plusplus}.
\end{minipage}   

\vspace{0.3cm}

    \begin{minipage}{0.68\textwidth}
    \includegraphics[width=\linewidth]{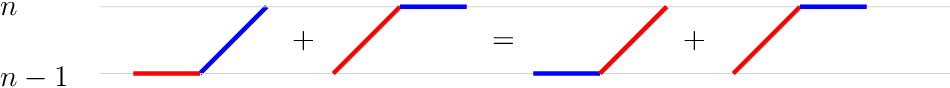}
\end{minipage}
\hfill
\begin{minipage}{0.28\textwidth}
    \small
    Eq. \eqref{zeroplus}.
\end{minipage}    

\vspace{0.3cm}

    \begin{minipage}{0.68\textwidth}
    \includegraphics[width=\linewidth]{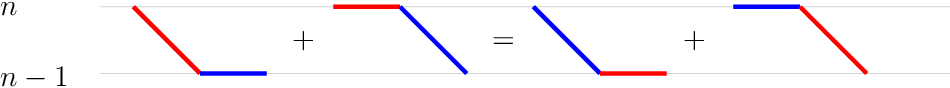}
\end{minipage}
\hfill
\begin{minipage}{0.28\textwidth}
    \small
    Eq. \eqref{zerominus}.
\end{minipage}  
    
\vspace{0.3cm}

        \begin{minipage}{0.68\textwidth}
    \includegraphics[width=\linewidth]{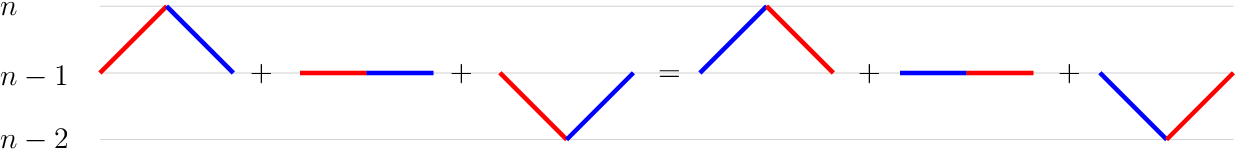}
\end{minipage}
\hfill
\begin{minipage}{0.28\textwidth}
    \small
    Eq. \eqref{plusminus}.
\end{minipage}      

 \caption{Graphical representation of the equations \eqref{minusminus}-\eqref{plusminus}. The face interaction tensor acts on the left-hand side of each graphical equation but is omitted from the diagrams to keep the notation  simple.}
    \label{figurelevels}
\end{figure}

\subsubsection{Solution}

We further explore the gauge invariance of
our ansatz 
\begin{eqnarray}
    &&Z_{s,s'} \to G Z_{s,s'} H^{-1}, \qquad 
    Z'_{s,s'} \to H Z'_{s,s'} G^{-1},
    \label{eq:gauge}\\
    &&L_{s,s'} \to L_{s,s'} H^{-1},\quad L'_{s,s'} \to L_{s,s'} G^{-1}, \qquad R_{s,s'} \to G R_{s,s'},\quad
R'_{s,s'}\to H R'_{s,s'}, \nonumber 
\end{eqnarray}
for arbitrary block-diagonal invertible matrices $G$ and $H$. We have essentially fixed $G$ and $H$ by numerically determining which matrix elements can be generally set to zero, 
which can be made identical, or which can be assigned trivial values ($\pm 1$), 
so that the resulting tensors have no remaining free parameters (i.e., the gauge freedom is fully fixed). 
In this way, we finally arrived at the ansatz that allowed for an explicit analytical solution of the boundary (\ref{ULZpboundary}-\ref{UZRpboundary})
and bulk-recurrence relations 
(\ref{minusminus}-\ref{plusminus}).
Specifically, for each level $n$, the number of identified unknown entries (variables) is listed in Table \ref{tab:levels}.

We find that, starting from level 5, both the number of unknown entries in the ansatz and the structure of the blocks in the tensors (see Figs.~\ref{figuremasks} and \ref{figuremasks2}) remain constant ($n$-independent).

\begin{table}[H]
\centering
\begin{tabular}{c|cccccc}
level $n$ & 0 & 1 & 2 & 3 & 4 & $\ge 5$ \\
\hline
entries &  12   &  15   &  91   &  107   &  102   & 97 
\end{tabular}
\caption{Number of unknown entries in the ansatz for each level in the direct sum decomposition (\ref{eq:directsum}). `Level $0$' refers to the unknown entries of the boundary tensors (\ref{BT}).}
\label{tab:levels}
\end{table}

We have obtained analytical expressions for $\LL, \LLp, \RR, \RRp$ and for $\ZZ$ and $\ZZp$ up to level $n=2$ for variable model parameters ($a,b,c,d,\beta,\gamma$). However, since the solution involves lengthy expressions we provided it externaly in~\cite{mathematicanotebook}, (Mathematica notebook \texttt{BoundaryNESS.nb}).

However, it was neither possible to solve the recurrence explicitly for any $n$ in the closed form, nor it was possible to obtain variable parameter solutions for $n\ge 3$.
Starting from level 3, we selected generic (arbitrary) values  for the bulk and boundary parameters and generated solutions of the system (\ref{minusminus}-\ref{plusminus}) within exact arithmetic. Levels up to $n\approx 40$ can be reached within minutes on a laptop.
We empirically found that the computational time required to reach level $n$ roughly scales as $t_{\rm compute} \sim O(n^{3.5})$, suggesting that an an explicit algorithm should exist for solving our recurrence.

The calculation was performed for several parameter choices; in particular, in the attached Mathematica notebook we report the results for
\begin{align}
    &a=\frac{11}{23}, &&b=\frac{19}{32}, &&c=\frac{23}{53}, &&d=\frac{31}{71}, &&\beta=\frac{30}{101}, &&\gamma=\frac{40}{49}, 
    \label{numericalparameters}
\end{align}
while the reader is free to modify the parameter values and re-iterate the solution.

It may be instructive to attempt to connect our patch matrix product ansatz 
to the patch state ansatz~\cite{prosen2016integrability}, 
which was derived for the NESS of the undeformed RCA54 model. 
However, in the limit $\gamma = \beta = 0$, our solution becomes singular, 
so a simple direct connection between the two results seems absent.

Furthermore, a rigorous proof of the following conjecture remains to be found:
 the recurrence system~\eqref{minusminus}--\eqref{plusminus}, with initial condition given by the boundary equations, always has a unique solution. Another obvious question to be explored is the construction of an exact solution for the other family of integrable boundaries identified in the undeformed RCA54 model~\cite{prosen2016integrability}, 
namely the so-called Bernoulli driving~\cite{buvca2021rule}.

\subsection{Digit complexity: empirical test of exact solvability of NESS}

The deformed RCA54 model appears to be a rather complex integrable model. Even though integrability structures emerge very clearly, their apparent complexity seems unusual compared to commonly discussed integrable models. One may even argue that
the results reported here are unnecessarily complicated, simply because we were unable to find an optimal gauge (e.g. (\ref{eq:gauge})) in which the results would look much nicer and simpler.

\begin{figure}[h]
    \centering
    \includegraphics[width=0.57\linewidth]{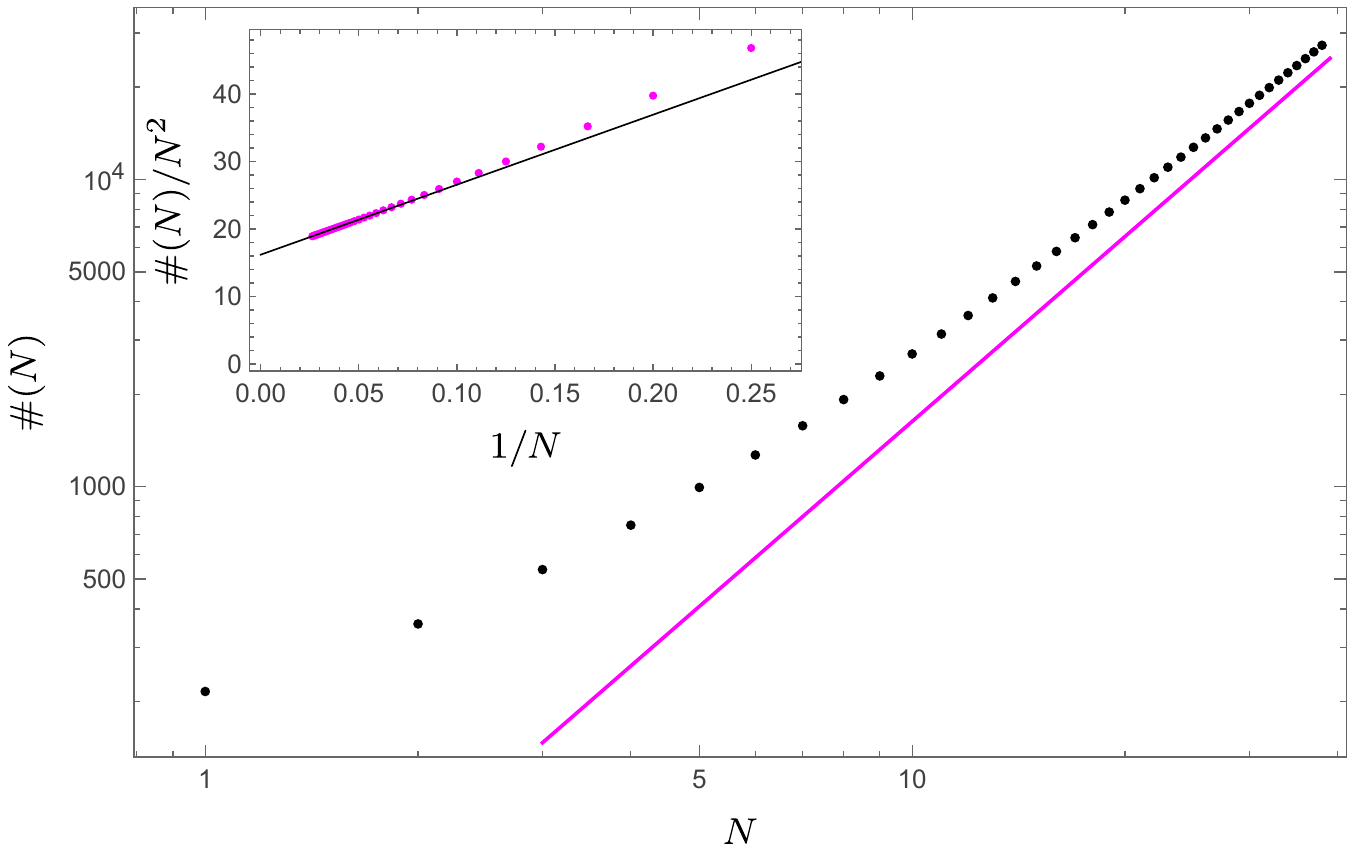}

    \includegraphics[width=0.36\linewidth]{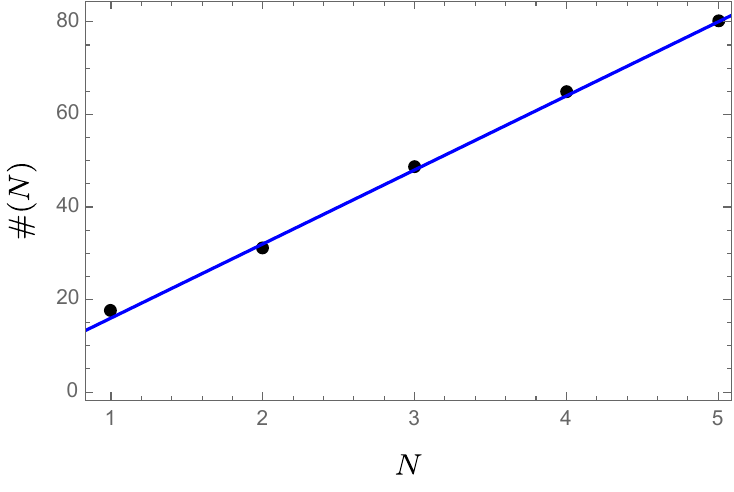}\quad\includegraphics[width=0.38\linewidth]{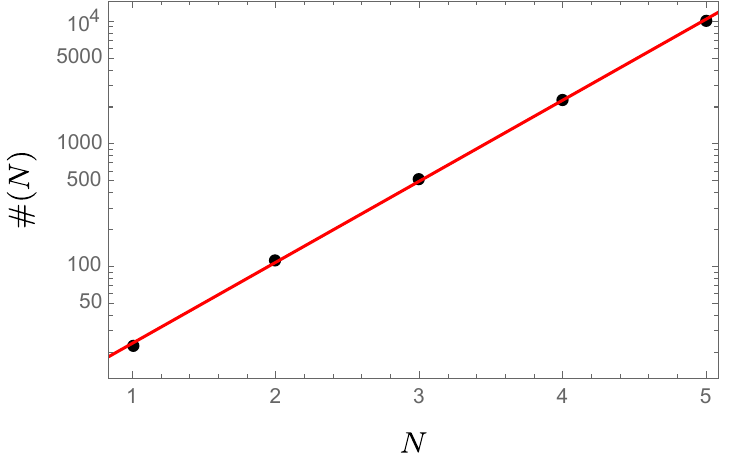}

    \caption{Top: The number of digits in the denominator of a rational gap probability $p_{00\cdots 0}$ in the NESS for a system of size $2N$, obtained from the exact arithmetic solution with parameters (\ref{numericalparameters}). The straight line indicates quadratic growth, $\propto N^2$. 
    Inset shows extrapolation of $N^2\#(N) \to {\rm const}$ as $1/N\to 0$.
    Bottom: Same quantity for the boundary-driven stochastic six-vertex model,
    which is defined on $2N+1$ sites with Markov matrix $\mathbb U=\mathbb U_{\rm e}\mathbb U_{\rm o}$,
    where $\mathbb U_{\rm e}=U^{\rm L}_1 U_{23}(p)\cdots U_{2N,2N+1}(p)$, 
    $\mathbb U_{\rm o}=U_{12}(q)\cdots U_{2N-1,2N}(q)U^{\rm R}_{2N+1}$, and $U_{i,i+1}(r) = 
    \frac{1}{1+r}\mathbb{1} + \frac{r}{1+r} P_{i,i+1}$, $U^{\rm L}={\tiny\begin{pmatrix}1-a & b\cr
    a & 1-b\end{pmatrix}}$, $U^{\rm R}={\tiny\begin{pmatrix}1-c & d\cr
    c & 1-d\end{pmatrix}}$. The model is integrable
    for $p=q$ and NESS can be expressed as a simple matrix product ansatz~\cite{vanicat2018integrable}, with the result displayed in bottom-left panel for $p=q=1/3$ with
    best-fit linear slope $16 N$ (blue), while it is
    almost surely non-integrable in the staggered case $p\neq q$, with the result displayed for
    $p=1/2$, $q=1/3$ with fitted exponential $5.2\exp(1.52 N)$ 
    (red). Boundary parameters in both cases are $a=1/4$, $b=3/5$, $c=3/4$, $d=2/7$.}
    \label{fig:arithmetic}
\end{figure}

In order to address this issue, and at the same time to have a simple criterion with which one can quantify potential integrability or exact solvability of a given empirically constructed object, such as a NESS, we propose here a notion of {\em digit complexity}.
Suppose that one can compute a gauge invariant physical observable, such as the gap-probability $p_{00\cdots 0}$ --- the probability of finding a completely empty configuration within a NESS --- using exact arithmetic for rational values of the model parameters $a,b,c,d, \beta, \gamma\in\mathbb Q$. Is there a universal behavior in the number of digits of the smallest denominator, $\#(N)$, required to express $p_{00\cdots 0} \in \mathbb Q$ as a function of the system size $N$?
This indicator is readily accessible within our Mathematica computation scheme for the stochastically deformed RCA54 and we plot it in Fig.~\ref{fig:arithmetic} for $N$ up to $80$. The result, which we have verified for several other choices of model parameters, indicates a simple universal asymptotic behavior
\begin{equation}
    \#(N) = O(N^2).
\end{equation}

We compare this with a more familiar exactly solvable boundary-driven system, specifically the boundary-driven (open) stochastic six-vertex model~\cite{vanicat2018integrable}. In that case, the number of digits clearly scales linearly, $\#(N) = O(N)$, as also follows from the known
oscillator algebra representation of the matrix product ansatz. This observation suggests that there is probably no simple closed form representation of
patch matrix ansatz tensors $\ZZ,\ZZp$.

However, if one breaks integrability (or exact solvability of the NESS) by introducing a staggered six-vertex parameter—making the hopping probabilities between sites $2n-1,2n$ and $2n,2n+1$ different—then we observe a clear indication of exponential scaling, $\log\#(N) = O(N)$, which we could define as a {\em positive} digit complexity $\kappa>0$, defined as $\kappa=\lim_{N\to\infty}\frac{\log\#(N)}{N}$.

In conclusion, although the number of digits can scale polynomially in various exactly solvable models, there may be a significant qualitative difference between models exhibiting linear versus superlinear (quadratic) growth, as in the deformed RCA54 case. It may therefore be of interest to further formalize this concept.

\label{digitcomplexity}

\section{Conclusion}

The deformed RCA54 appears to be an unusual and fascinating integrable model. In several respects, it behaves quite differently from more familiar integrable vertex models, such as the six-vertex and eight-vertex models, as well as from related integrable quantum spin chains. The results presented here, although partly based on well-supported conjectures, call for further investigation of this model and perhaps even motivate a systematic approach to the general classification of integrable interaction-round-a-face (IRF) quantum and/or stochastic many-body dynamical systems. As discussed  in~\cite{Carr1,Carr2}, the physical phenomenology of more general IRF QCA dynamics can be very rich. For instance, in Ref.~\cite{Carr2} another class of
exactly solvable QCA has been discussed, namely the so-called Goldilocks rules, which can be related to free fermions.

In this work, we have discussed two physical realizations of the model: (i) the commuting transfer matrix and conserved local charges of the quantum deformation of RCA54 with periodic boundary conditions (or on an infinite lattice), and (ii) the exact solution for the non-equilibrium steady state of the stochastic deformation of RCA54 with stochastic boundaries. Both settings invite substantial further generalization.

In the quantum case (i), one may ask whether the Lax operators admit further simplifications or factorizations. Another natural question is whether the present deformation is already the most general one and, more broadly, how one might achieve a complete classification of integrable IRF models of qubits ($d=2$). It would furthermore be interesting to develop an IRF analog of the reflection algebra and to classify the corresponding integrable unitary boundary processes.

In the stochastic case (ii), we noted empirical evidence suggesting that the full Markov generator of the time evolution, and not merely its fixed point (the NESS), is likely integrable, as indicated by the complex spacing ratio criterion~\cite{sa2020complex}. Devising a framework for the full Bethe-ansatz diagonalization of this Markov chain therefore remains an interesting open problem. Moreover, even in the absence of stochastic boundaries, the model can be checked numerically to exhibit KPZ fluctuations---for instance, by interpreting the number of lines crossed in $t$ time steps of a trajectory in Fig.~\ref{figuretraj} as a height function---with growth exponent $1/3$ for the asymmetric deformation $\beta\neq\gamma$, and Edwards--Wilkinson scaling with growth exponent $1/4$ for the symmetric deformation $\beta=\gamma$. This is naturally consistent with the observation that the stochastically deformed RCA54 corresponds to a discretization of the $t$-PNG model~\cite{borodin2023}, although further work is needed to firmly establish these observations and connections.

Finally, the deformed RCA54 remains integrable even for completely arbitrary complex deformation parameters, beyond the unitary and stochastic regimes. In this broader setting, it may eventually find interesting applications in the context of non-Hermitian physics.

\ack{We would like to thank M. P. T. Costa, M. de Leeuw, D. Fioravanti, J. Links, K. Mallick, L. Piroli, B. Pozsgay, A. L. Retore, R. Sharipov and E. Vernier for useful discussions.}

\funding{We acknowledge funding from the European Union HORIZON-CL4-2022-QUANTUM-02-SGA through PASQuanS2.1 (Grant Agreement No. 101113690), European Research Council (ERC) through Advanced grant QUEST (Grant Agreement No. 101096208), as well as the Slovenian Research and Innovation agency (ARIS) through the Program P1-0402 and Grant N1-0368.}
\begin{appendix}
\numberwithin{equation}{section}

\newpage
\section{Masks}
\label{masksappendix}
We present here the block structure of the solution for $\ZZ$, $\ZZp$, $\LL$, $\LL'$, $\RR$, and $\RR'$, using the numerical values of the parameters in (\ref{numericalparameters}).

Black and gray squares indicate the non-zero entries. Starting from level 5, the block structure of the ansatz is identical for all subsequent levels. For the labeling of the levels, we refer to Fig. \ref{figureZansatz}.
\begin{figure}[H]
\centering    \includegraphics[width=1\textwidth]{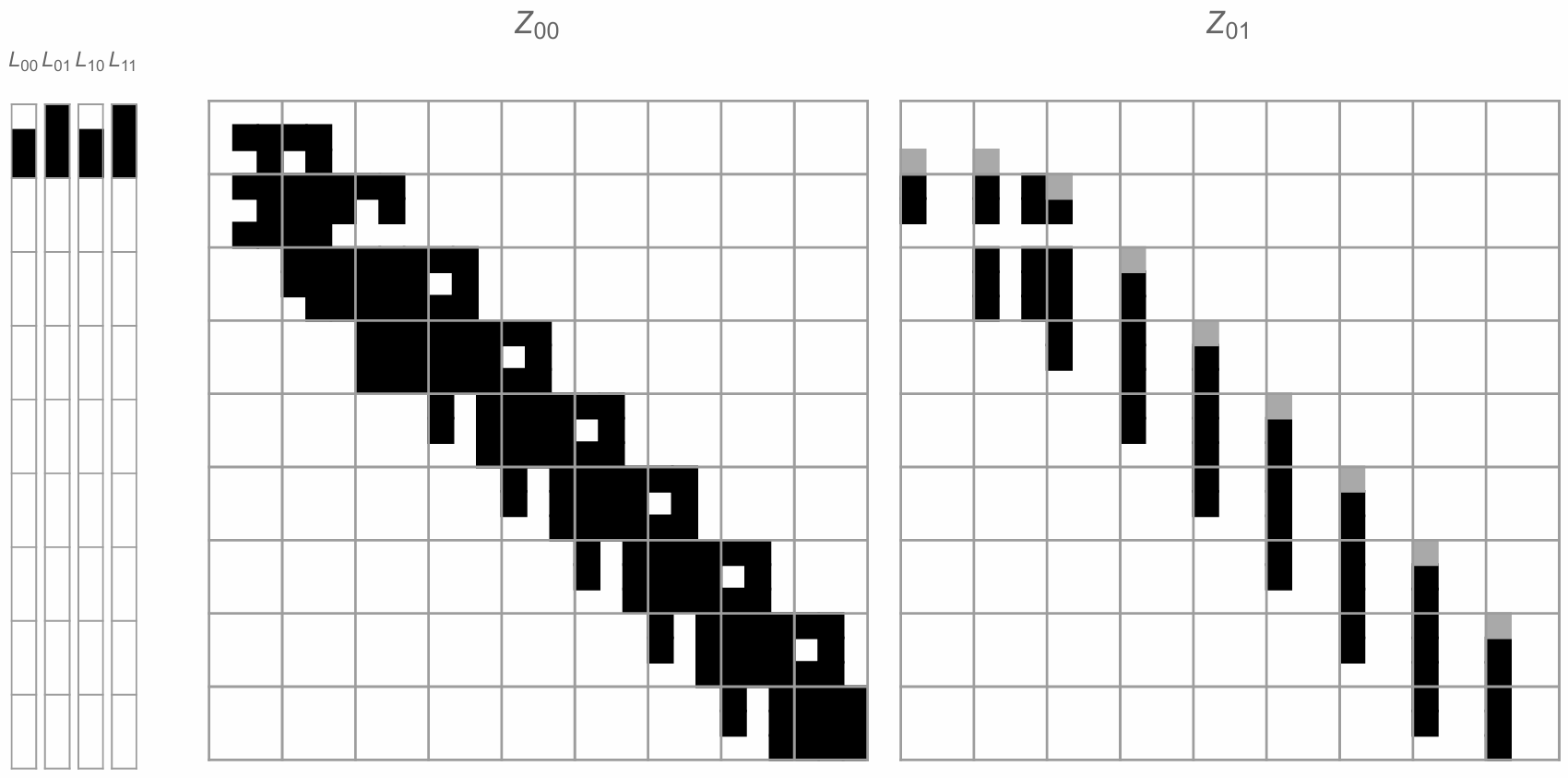}
\centering    \includegraphics[width=1.\textwidth]{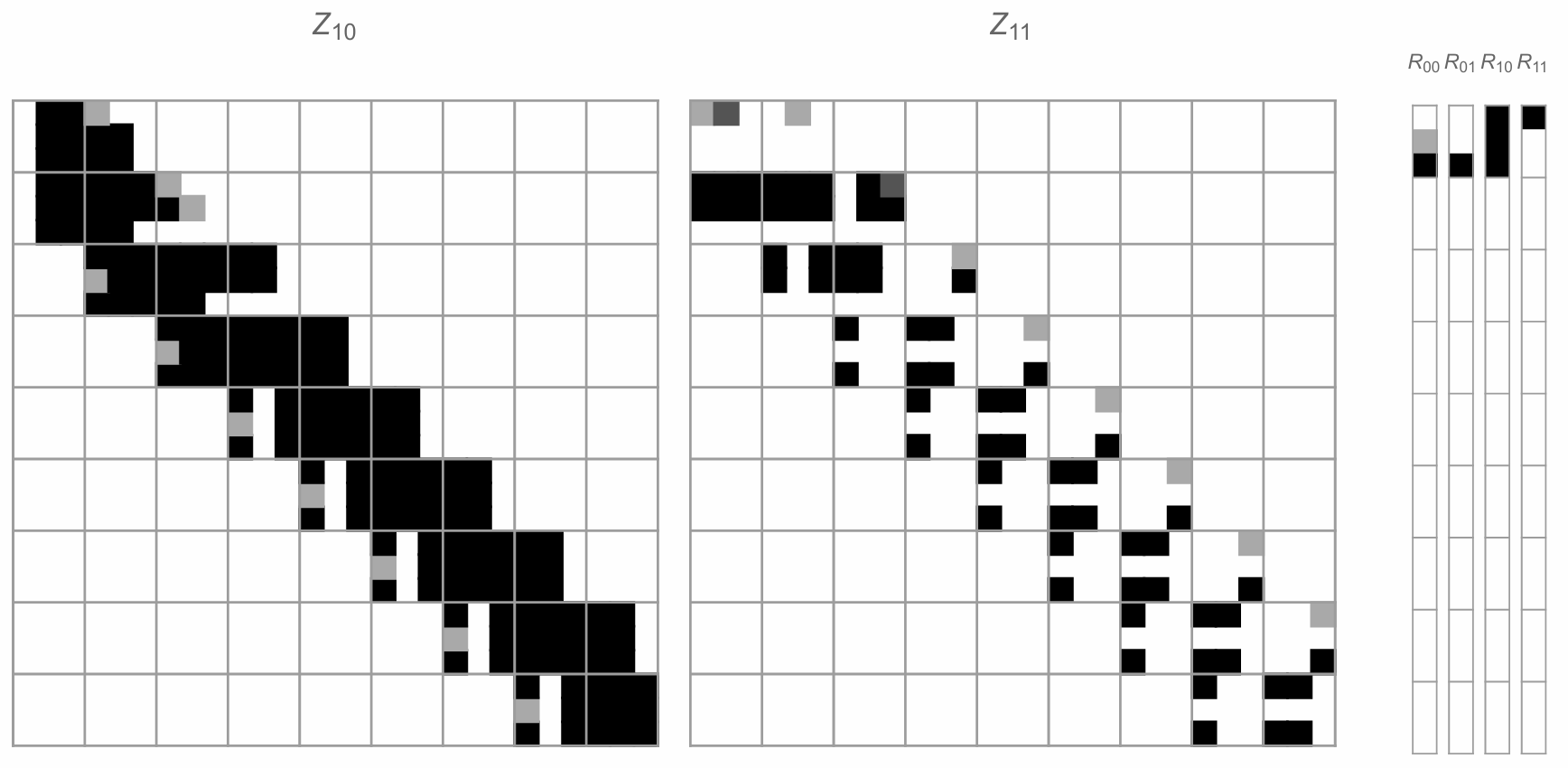}
    \caption{Non-zero entries in the infinite-dimensional matrix $\ZZ$ up to level 9, as well as in the boundary vectors $\LL$ and $\RR$. Grey boxes correspond to $1$, dark grey boxes to $-1$, and black boxes indicate entries different from $0$.}
    \label{figuremasks}
\end{figure}

\begin{figure}[H]
\centering    \includegraphics[width=1\textwidth]{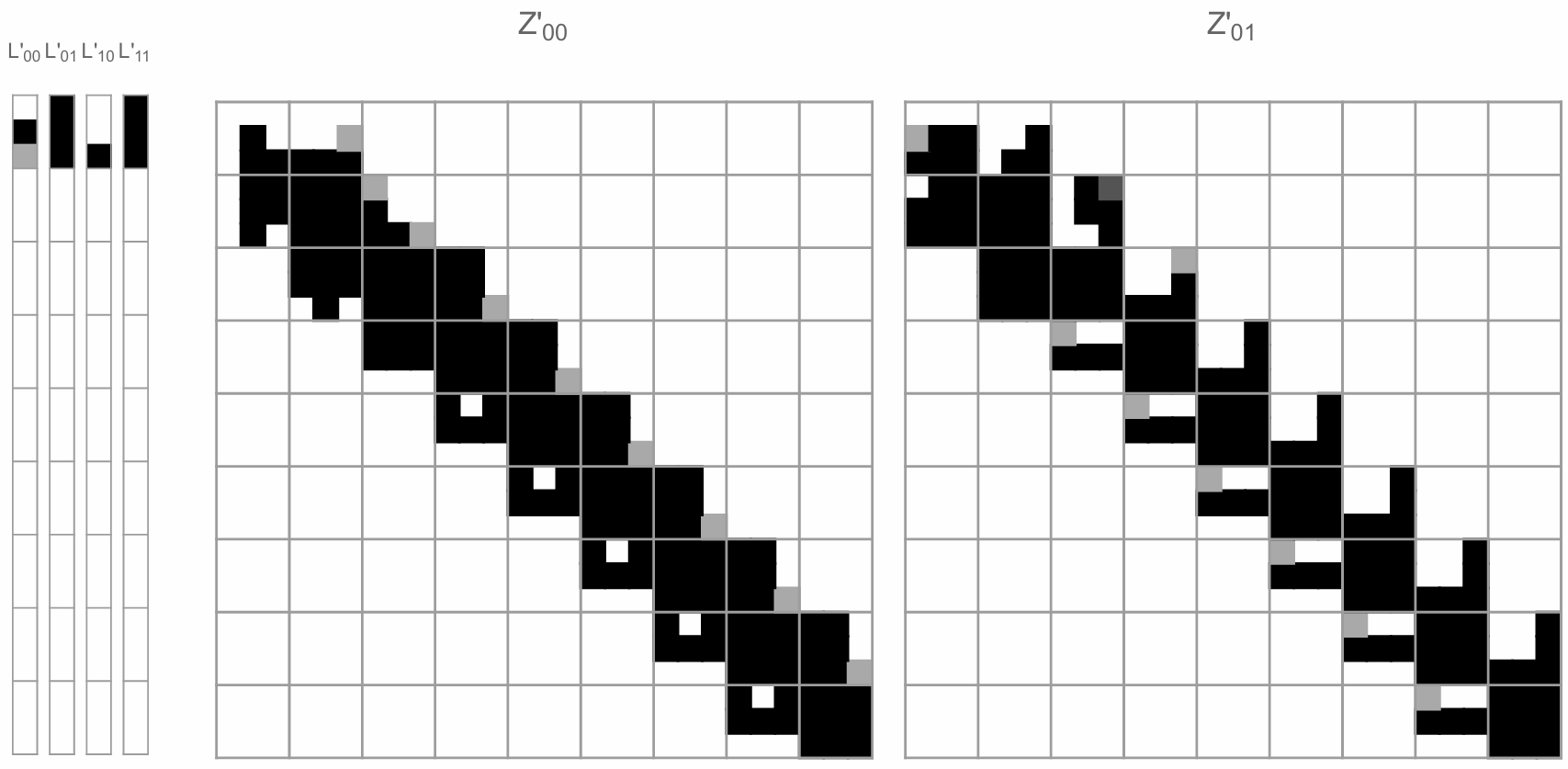}
\centering    \includegraphics[width=1.\textwidth]{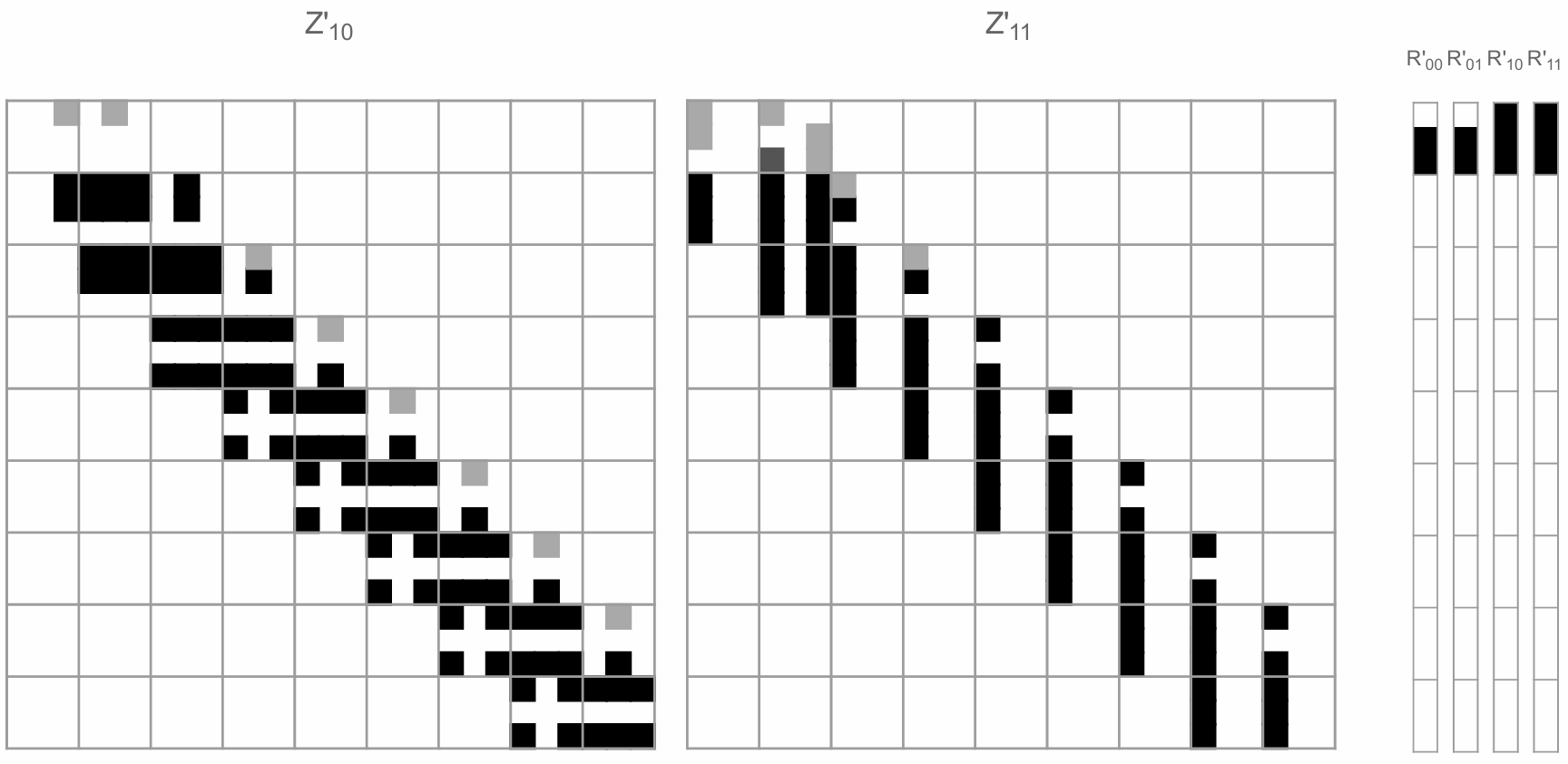}
    \caption{Non-zero entries in the infinite-dimensional matrix $\ZZp$ up to level 9, as well as in the boundary vectors $\LL$ and $\RR$. Grey boxes correspond to $1$, dark grey boxes to $-1$, and black boxes indicate entries different from $0$.}
    \label{figuremasks2}
\end{figure}

\section{Walk through the attached Mathematica notebooks}
\label{notebookappendix}
In \cite{mathematicanotebook}, we provide two Mathematica notebooks: \texttt{BulkIntegrability.nb}, which covers the results on bulk integrability, and \texttt{BoundaryNESS.nb}, which addresses the boundary part. In this appendix, we explain the different sections of the notebooks, with some comments within the notebooks serving as an additional guide.

\subsection{Bulk: \texttt{BulkIntegrability.nb}}
In this notebook, we provide the expressions for the conserved charges $Q_6, Q_{10}, Q_{14}$ for arbitrary values of the deformation parameters $\alpha, \beta, \gamma, \delta$, and use them to construct the perturbative expression of the Lax operator \eqref{ansatzL1} up to order $u^3$. Additionally, by fixing the parameters to the numerical values in \eqref{numericalchoice}, we present the perturbative expansion up to order $u^{35}$, as well as its full analytical expression.

We also include a guide to the different sections of the notebook.

\begin{itemize}
    \item[1.] \textit{Fix dimensions of the circuit} The input of this section is the dimension (length) of the circuit $L$. Since we want to work with a translationally invariant model, we consider the single-glued picture, as explained in Point 2 of Sec. \ref{perturbativesketch}. Therefore, $L=\frac{N}{2}$.
    \item[2.] \textit{Definition} We provide the necessary definition to embed the local density charges into extensive charges.
    \item[3.] \textit{Dynamics} We compute the evolution operator $\mathbb{U}$ as defined in \eqref{Ucircuit}.
    \item[4.] \textit{Conserved charges $q_3$, $q_5$ and $q_{7}$} We provide the analytical expressions of the charges $\mathcal{Q}_3$, $\mathcal{Q}_5$, and $\mathcal{Q}_7$, as given in the main text in \eqref{Q3fromh}, \eqref{Q5standardspace}, and \eqref{Q7standardspace}. These depend on $\rom{h}$, $\tilde{\rom{h}}$, and $\tilde{\tilde{\rom{h}}}$.
    
    In the notebook, the densities of these charges are denoted by $\text{q}3$, $\text{q}5$, and $\text{q}7$. Correspondingly, the symbols in the main text are mapped in the notebook as $\rom{h} \to \text{q}3$, $\tilde{\rom{h}} \to \text{htilde}$, and $\tilde{\tilde{\rom{h}}} \to \text{hhtilde}$.
    
    Note that $\tilde{\rom{h}}$ and $\tilde{\tilde{\rom{h}}}$ depend on some unfixed parameters $\eta_i$ and $\eta2_i$. The reader can verify that the commutation relations between the charges and $\mathbb{U}$ hold for any choice of these parameters. They merely correspond to a gauge freedom and can be eliminated by enforcing orthonormality between the charges. We have chosen not to fix these parameters here in order to demonstrate that these charges are equivalent to those obtained in the second part of the notebook, where the parameters $\alpha, \beta, \gamma, \delta$ are fixed as in \eqref{numericalchoice}. We make this comparison in Sec. "Lax operator analytical expression for $\alpha=1/7;\, \beta=1/2;\,\gamma=1/8;\,\delta=3/11$".
    
    We also provide the perturbative expression of the Lax operator given in \eqref{ansatzL}, denoted as \\$\text{Lcheckpert[u]}$.
    \item[5.] \textit{Lax operator analytical expression for $\alpha=1/7;\, \beta=1/2;\,\gamma=1/8;\,\delta=3/11$} We denote by Lpertnum[u] and Lchecknum[u] the perturbative expressions of the Lax operators $\mathcal{L}(u)$ and $\check{\mathcal{L}}(u)$ up to order $u^{35}$. We have verified that these expressions are consistent with the charges given in Sec. “Conserved charges $q_3$, $q_5$, $q_{10}$.” For instance, we computed $\tilde{\rom{h}}$ and $\tilde{\tilde{\rom{h}}}$ from \eqref{secondderivative} and \eqref{thirdderivative}, and checked that, for specific choices of $\eta_i$ and $\eta2_i$, the expressions agree.
    
    Lax[u] represents the ansatz for the non-zero entries of the Lax operator. $\check{\mathcal{L}}(u)$ is shown in Fig. \ref{nonzeroLax}.
    
    We also provide the equation for the entry $f_{120}(u)$ of this ansatz, given in \eqref{f120entries}, along with the explicit expressions of the polynomials $W_2(u)$, $W_4(u)$, $F_4(u)$, and $G_4(u)$. Since $f_{120}$ is the only entry where a square root appears, all other entries are defined as functions of $f_{120}$.
\end{itemize}

\subsection{Boundary: \texttt{BoundaryNESS.nb}}
In this notebook, we provide a solver for the boundary and bulk equations \eqref{ULZpboundary}–\eqref{UZZbulk}. In particular, the reader can obtain solutions for the boundary vectors $\LL, \RR, \LLp, \RRp$ and the bulk tensors $\ZZ, \ZZp$ up to a fixed chosen level, for a given numerical choice of the boundary and bulk parameters $a$, $b$, $c$, $d$, $\beta$, $\gamma$. Furthermore, we provide analytical solutions for $\LL, \RR, \LLp, \RRp$ and for $\ZZ, \ZZp$ up to level 2.

We explain the different sections of the notebook.

\begin{itemize}
    \item [1.] \textit{Input: lmax and $a$, $b$, $c$, $d$, $\beta$, $\gamma$} The input of this section is lmax, which corresponds to the maximal requested level in the ansatz for $\ZZ$ and $\ZZp$, as well as the boundary and bulk parameters $a$, $b$, $c$, $d$, $\beta$, $\gamma$. These parameters are set to the values in \eqref{numericalparameters}, but they can be changed to arbitrary (rational) values.
    \item[2.] \textit{Ansatz} We provide the ansatz for the boundary vectors $\LL, \LLp, \RR, \RRp$ and the bulk tensors $\ZZ,\ZZp$.

\begin{itemize}
    \item[2a.]{Boundary:} We refer to La[$s,s'$], Lb[$s,s'$], Ra[$s,s'$], Rb[$s,s'$] to denote $L_{s,s'},\,{L'}_{s,s'},\,R_{s,s'},\,{R'}_{s,s'}$, where $s,s'$ identify the components. We fix the ansatz for these vectors so that they solve \eqref{URboundary} and \eqref{ULppboundary}. To determine them completely, we also need to solve Eqs. \eqref{ULZpboundary} and \eqref{UZRpboundary}. The entries of the vectors are labeled as x[0,i], where $i$ indexes the different components. We note that, while these vectors require only level-1 components, we label them with 0 for simplicity, in order to distinguish them from the boundary vectors.
    \item[2b.] {Bulk:} We denote the $3 \times 3$ block structures as Za[$s,s',m,n$] (for $Z_{s,s'}^{(n,m)}$) and Zb[$s,s',m,n$] (for ${Z'}_{s,s'}^{(n,m)}$), where $s,s'$ identify the components, and $m = 0, +1, -1$ corresponds, respectively, to the diagonal block, the block above ($+1$), or the block below ($-1$) and $n$ is the level, as depicted in Fig. \ref{figureZansatz}.
    
    By running the various commands in the code, one obtains the ansatz for the masks provided in Appendix \ref{masksappendix}. The code structure (the comments in the code provide further guidance) accounts for the fact that the ansatz of the blocks from level $5$ onward must be the same. To keep track of the entries in the ansatz, we use the label $x[n, i]$ where $n$ is the level and $i$ indexes the different entries in $Z, \,Z'$.
\end{itemize}
\item[3.] \textit{Equations} We write the dynamical evolution operator $\mathbb{U}$ as U (see \eqref{propagatorU}) and UL, UR are defined in \eqref{conditionaldriving}. The bulk equations bulkeq[$n$] are given in \eqref{minusminus}–\eqref{plusminus}, while the boundary equations are separated as follows: boundaryR1 and boundaryL1 correspond to \eqref{URboundary} and \eqref{ULppboundary}, and boundaryR2, boundaryR3, boundaryL2, and boundaryL3 correspond to \eqref{UZRpboundary} and \eqref{ULZpboundary} for levels 1 and 2. As mentioned, the ansatz chosen for the boundary vectors solve boundaryR1 and boundaryL1.
\item[4.] \textit{Solutions} Here, we provide a strategy to solve the bulk and boundary equations. We proceed level by level. We start by solving the equations eqboundR2 and eqboundL2. Solving these equations yields several solutions; however, most of them have some entries $x[1,i]$ equal to zero. Since, in our ansatz, all remaining entries must be non-zero, we select the solutions for which all entries are non-zero. There is only one solution of this type. We call it sol1.

Using a similar strategy, we then substitute the obtained solution into eqboundR3, eqboundL3, and eqbound[2] and we solve them for the entries of level 2 (sol2). We continue this procedure, solving the equations level by level up to lmax. We refer to them as soltab[n], with $n=3,4,\dots, \text{lmax}.$

With this approach, we complete the solution of our problem and provide a method to obtain an explicit expression for the NESS of the deformed Rule 54 model with conditional boundary terms.

In the notebook, we provide an explicit solution for the numerical values \eqref{numericalparameters} and lmax=15. The reader can easily run the code for any choice of parameters. In the solution, we present the 3x3 blocks depicted in Figs. \ref{figuremasks} and \ref{figuremasks2} in this form
\begin{center}
$Z[n,s,s']: \,\,\,\,\,Z_{s,s'}^{(n,+)},\,\,\,\,\, Z_{s,s'}^{(n,0)}, \,\,\,\,\,Z_{s,s'}^{(n,-)}$
\\
$Zp[n,s,s']: \,\,\,\,\,Z_{s,s'}^{\prime\,(n,+)}, \,\,\,\,\,Z_{s,s'}^{\prime\,(n,0)}, \,\,\,\,\,Z_{s,s'}^{\prime\,(n,-)}$
\end{center}
where the superscript index $\pm,0$ refers to the convention of Fig. \ref{figureZansatz}.
\item[5.] \textit{Extra Simplifications} We present further analytical relations among the entries that hold at all levels (higher than 5 or 6) and for general parameter values. However, some of these relations are nonlinear. We found that enforcing them in the ansatz from the outset increases the computational cost of solving level by level, with scaling growing as $\sim n^{4.7}$ instead of $\sim n^{3.5}$. For this reason, we chose not to impose them initially.

The subsections “same level”, “one level less,” and “two levels” classify the types of relations considered. These relations either involve entries at the same level (valid from level 5) or take the form $x[i,n] = f(x[j,n-1])$ or $x[i,n] = f(x[j,n-1], x[k,n])$ (valid from level 6), where $f$ denotes a functional dependence of one entry on others.
\item[6.] \textit{Analytic up to level 3} Here, we present the analytical solutions for the entries up to level 3, for variable values of the parameters $a$, $b$, $c$, $d$, $\beta$, and $\gamma$. In particular, ALLVARpartialnew and ALLVAR2new contain 100 entries (including those at the boundaries, level 1, and part of level 2). As indicated in Table \ref{tab:levels}, the total number of unknown entries is 118. The remaining 18 entries are particularly lengthy, and are therefore provided in the separate text file ``\texttt{merged\_all.txt}'' of \cite{mathematicanotebook}.
\end{itemize}

\end{appendix}

\bibliographystyle{iopart-num}
\bibliography{SciPost_Example_BiBTeX_File}

\end{document}